\documentclass[bibyear]{aa} 

\usepackage{natbib}
\bibpunct{(}{)}{;}{a}{}{,} 
\usepackage{graphicx}
\usepackage{txfonts}
\usepackage{float}
\usepackage{xcolor}
\usepackage{siunitx}
\usepackage{hyperref}
\begin{document}

\title{Testing General Relativity: new  measurements of gravitational redshift in galaxy clusters}

\author{D. Rosselli \inst{1}\fnmsep\inst{2}\fnmsep\thanks{Corresponding author:
    Damiano Rosselli\\ \email{damiano.rosselli@studio.unibo.it}}
  \and F. Marulli\inst{1}\fnmsep \inst{3}\fnmsep \inst{4}
  \and A. Veropalumbo\inst{5}\fnmsep \inst{6}
  \and A. Cimatti\inst{1}\fnmsep \inst{7}
  \and L. Moscardini\inst{1}\fnmsep \inst{3}\fnmsep \inst{4} }

\institute{Dipartimento di Fisica e Astronomia ``Augusto Righi'' -
  Alma Mater Studiorum Università di Bologna, via Piero Gobetti
  93/2, I-40129 Bologna, Italy
  \and Aix Marseille Univ, CNRS/IN2P3, CPPM, Marseille, France
  \and INAF - Osservatorio di Astrofisica e Scienza dello Spazio di
  Bologna, via Piero Gobetti 93/3, I-40129 Bologna, Italy
  \and INFN - Sezione di Bologna, viale Berti Pichat 6/2, I-40127
  Bologna, Italy
  \and Dipartimento di Fisica, Universit\`{a} degli Studi Roma Tre,
  via della Vasca Navale 84, I-00146 Rome, Italy
  \and INFN - Sezione di Roma Tre, via della Vasca Navale 84,
  I-00146 Rome, Italy
  \and INAF - Osservatorio Astrofisico di Arcetri, Largo E. Fermi 5, I-50125 Firenze, Italy
  }


\abstract
{ The peculiar velocity distribution of cluster member
  galaxies provides a powerful tool to directly investigate the
  gravitational potentials within galaxy clusters and to test the
  gravity theory on megaparsec scales. }
{ We exploit spectroscopic galaxy and galaxy cluster
  samples extracted from the latest releases of the Sloan
    Digital Sky Survey (SDSS) to derive new constraints on the
  gravity theory.}
{  We consider a spectroscopic sample of $\num{3058}$
  galaxy clusters, with a maximum redshift of $0.5$ and masses between
  $10^{14} - 10^{15}$ M$_{\odot}$.  We analyse the velocity
  distribution of the cluster member galaxies to make new measurements
  of the gravitational redshift effect inside galaxy clusters. We
  accurately estimate the cluster centres, computing them
  as the average of angular positions and redshifts of the closest
  galaxies to the brightest cluster galaxies. We find that this centre
  definition provides a better estimation of the centre of the cluster
  gravitational potential wells, relative to simply assuming the
  brightest cluster galaxies as the cluster centres, as done in the
  past literature.  We compare our measurements with the theoretical
  predictions of three different gravity theories: general relativity
  (GR), the $f(R)$ model, and the Dvali–Gabadadze–Porrati (DGP)
  model. A new statistical procedure is used to fit the measured
  gravitational redshift signal and thus to discriminate among the
  considered gravity theories. Finally, we investigate the systematic
  uncertainties possibly affecting the analysis.}
{We clearly detect the gravitational redshift effect in
  the exploited cluster member catalogue. We recover an integrated
  gravitational redshift signal of $-11.4 \pm 3.3$ km s$^{-1}$, which
  is in agreement, within the errors, with past literature works.}
{ Overall, our results are consistent with both GR and DGP
  predictions, while they are in marginal disagreement with the
  predictions of the considered $f(R)$ strong field model.}


\maketitle
%

\section{Introduction}

The $\Lambda$-cold dark matter ($\Lambda$CDM) model is currently
considered the standard cosmological framework and provides a
satisfactory description of the Universe on the largest scales
\citep{amendola2018, planck2018}. Einstein's theory of general
relativity (GR) is the foundation of all the equations that describe
how the Universe evolves and the formation of the cosmic structures we
can observe today. During the past years, GR has been systematically
tested both on small and large cosmological scales \citep[see
  e.g.][and references therein]{beutler2014, moresco2017}, though
current measurements are not accurate enough to discriminate among the
many alternative theories of gravity that were proposed to explain the
accelerated expansion of the Universe and the growth of cosmic
structures.

Clusters of galaxies are the most massive virialized structures in the
Universe. Thanks to their high masses and deep gravitational
potentials, it is possible to test GR on the scales of these large
structures by measuring the gravitational redshift through the
peculiar velocity distribution of cluster member galaxies
\citep{cappi,Kim_2004}.

The first detection of the gravitational redshift effect in galaxy
clusters was made by \citet{Wojtak_2011} using data from the seventh
data release (DR7) of the Sloan Digital Sky Survey
\citep[SDSS,][]{Abazajian_2009} and the Gaussian Mixture Brightest
Cluster Galaxy (GMBCG) sample \citep{Hao_2010}. \citet{Wojtak_2011}
measured the gravitational redshift signal up to $6$ Mpc from the
cluster centre, which was assumed to be coincident with the brightest
cluster galaxy (BCG) position. Their measurements were in agreement
with both GR and $f(R)$ theories. Similar analyses have been performed
by \citet{Jimeno_2015} and \citet{Sadeh_2015} using SDSS DR10 data
\citep{Ahn_2014}. These authors, differently from \citet{Wojtak_2011},
included in their theoretical model the effects described in
\citet{kaiser2013}.  In particular, \citet{Jimeno_2015} measured the
gravitational redshift signal up to $7$ Mpc from the cluster centre,
analysing three different cluster catalogues, that is the GMBCG, the
sample described in \citet{Wen_2012} (WHL12), and the RedMaPPer
cluster sample \citep{Rykoff2014}. The gravitational redshift effect
was measured both as a function of the distance from the cluster
centre, and as a function of the cluster masses. \citet{Jimeno_2015}
detected a significant signal in the GMBCG and RedMaPPer samples,
while the measurements in the WHL12 sample were not in agreement with
theoretical expectations.  The latest attempt was carried out by
\citet{gravfromspider}, who analysed the SPectroscopic IDentification
of ERosita Sources \citep[SPIDERS,][]{spidercat} survey. In
particular, they considered three different definitions of the cluster
centre, that is the BCG position, the redMaPPer identified central
galaxies, and the peak of the X-ray emission. With all the three
centre definitions, they obtained a clear detection of the
gravitational redshift, but their results could not discriminate
between GR and $f(R)$ predictions.

Our work aims at updating and improving these past analyses exploiting
the new galaxy data released by SDSS DR16 \citep{DR16} and the new
galaxy cluster sample provided by \cite{wen15} (WH15). We refine the
measurement method and the theoretical model to improve the accuracy
of the analysis. Thanks to these improvements we are able to reduce
the measurement errors by about $\num{30} \%$ with respect to the
works of \citet{Sadeh_2015} and \citet{gravfromspider}, up to a
distance of almost 3 Mpc from the cluster centres. The huge number of
measured redshifts inside our sample allows us to perform an accurate
Bayesian analysis, imposing new constraints on GR on megaparsec
scales. 

The paper is organised as follows. In Sec. \ref{data} we introduce the
analysed cluster catalogue and the SDSS DR16 galaxy sample, while in
Sec. \ref{searchclustmemb} we describe the new cluster member
catalogue we have constructed for this analysis. In Sec. \ref{model}
we present the theoretical predictions on the galaxy line-of-sight
velocity distribution offsets as a function of the distance from the
cluster centre in three different gravity theories. The method used to
measure this statistic from the observed galaxy redshifts is described
in Sec. \ref{result}. In Sec. \ref{measures} we present the main
results of our work. In Sec. \ref{conclusions} we conclude with
closing remarks and future prospects. Finally, in Appendix
\ref{appendix_systematics} we describe the analysis of the systematic
uncertainties affecting our measurements.

In this work all the cosmological calculations are performed assuming
a flat $\Lambda$CDM model, with $\Omega_m = 0.3153 $ and $H_0 = 67.36$
km s$^{-1}$ Mpc$^{-1}$ \citep[][Paper VI: Table $2$,
  TT,TE,EE+lowE+lensing,]{planck2018}. The whole cosmological analysis
has been performed with the \texttt{CosmoBolognaLib}
\citep[CBL,][]{CBL}, a large set of {\em free software} C++/Python
libraries that provides an efficient numerical environment for
statistical investigations of the large-scale structure of the
Universe. The new likelihood functions for fitting the velocity
distributions and computing GR and the alternative gravity theory
predictions, will be released in the forthcoming public version of the
CBL.

\section{Data}\label{data}

\subsection{The cluster sample }\label{whlclustcat}

In this work we exploit the galaxy cluster sample described in
\citet{wen15}, which is an updated version of the WHL12 cluster
catalogue.  The WHL12 sample was built using the SDSS-III photometric
data \citep[SDSS DR8,][]{Aihara_2011}. The method used to identify the
galaxy clusters is based on a \textit{friend-of-friend} algorithm. In
practice, a cluster is identified if more than eight member galaxies,
with an \textit{r}-band absolute magnitude smaller than $-21$, are
found within a radius of $0.5$ Mpc and within a photometric redshift
range of $0.04(1+z)$. After that, the BCG is recognised among the
cluster members and it is taken as the cluster centre. WHL12
calculated the total luminosity within a radius of $1$ Mpc,
$L_{1\text{Mpc}}$, then by using a scaling relation between
$L_{1\text{Mpc}}$ and the cluster virial radius $r_{200}$, they
computed $r_{200}$. The total luminosity within the $r_{200}$ radius
and the cluster richness were eventually computed. The optical
richness, R$_{\text{L}200}$, was used as a proxy for the cluster mass,
$M_{200}$, within $r_{200}$.

In the WH15 catalogue the cluster masses have been
re-calibrated. Specifically, WH15 exploited new cluster mass
estimation from X-ray and Sunyaev–Zeldovich effect measurements to
re-calibrate the richness-mass relation within the redshift range
$0.05 < z < 0.75$.  The calibrated relation can be expressed as
follows:
\begin{equation}\label{richness-mass}
  \log(M_{500})= 14 + (1.08 \pm 0.02)\log(\text{R}_{\text{L}500}) -
  (1.37 \pm 0.02)\,,
\end{equation}
where $M_{500}$ and $\text{R}_{\text{L}500}$ are the mass and the
optical richness within $r_{500}$\footnote{$r_{500}$ is the radius
where the cluster density is equal to 500 times the Universe critical
density.}, respectively.  By using the spectroscopic data of the SDSS
DR12 \citep{Alam_2015}, the authors also extended the number of
clusters with spectroscopic redshifts.

The final sample includes the data of BCG angular positions and
redshifts of $\num{132684}$ clusters within the redshift range $0.05
\leq z \leq 0.8$. The identified clusters have an average redshift of
$\langle z \rangle = 0.37$, an average mass of $\langle M_{500}
\rangle = 1.4 \times 10^{14}$ M$_{\odot}$, and an average radius of
$\langle r_{500} \rangle = 0.67$ Mpc.  The authors claimed that this
sample is almost complete in the redshift range $0.05 \leq z < 0.42$
and for masses above $10^{14}$ M$_{\odot}$.

\subsection{The spectroscopic galaxy samples}\label{galsample}

We exploit the galaxy coordinates and spectroscopic redshifts derived
from SDSS DR16 \citep{DR16}. Specifically, we analyse the data
collected by the Baryon Oscillation Spectroscopic Survey
\citep[BOSS][]{boss}, the Extended Baryon Oscillation Spectroscopic
Survey \citep[eBOSS][]{eboss} and the Legacy Survey obtained as part
of the SDSS-I and SDSS-II programs \citep{legacy}. Although the
spectroscopic data and the sky coverage of the galaxy samples have
remained unchanged during the past years, the imaging and the
spectroscopic pipelines have been improved in subsequent SDSS data
releases. Therefore, in this work we use the data of the latest
release.

The Legacy Survey covers a total sky area of $8032$ deg$^2$ and it is
composed of two galaxy samples: the \textit{Main sample}, a magnitude
limited sample of galaxies with a mean redshift of $z \simeq 0.1$
\citep{strauss2002} and the \textit{Luminous Red Galaxies} (LRG)
sample, a volume-limited sample up to $z \simeq 0.4$
\citep{Eisenstein_2001}.

Within the Legacy Survey, we select the galaxies with the most
reliable spectra and the lowest redshift errors. Specifically, we
select the objects in the catalogue which have the following
flags\footnote{The official SDSS DR16 web site
(https://www.sdss.org/dr16/) provides a detailed description of the
flags in the spectroscopic catalogues.}: \textsc{SPECPRIMARY} equal to
$1$, \textsc{CLASS} ``Galaxy'', \textsc{ZWARNING} equal to $0$, $4$ or
$16$, \textsc{ZERR} less than $6 \times 10^{-4}$ and \textsc{Z}
between $0.05$ and $0.75$ included. These selections are applied to
avoid multiple entries of the same object in the final catalogue, and
to include only those galaxies with reliable spectroscopic redshift
measurements. Moreover, we select the redshift range where the
richness-mass relation of the cluster sample has been calibrated.  We
find almost $\num{760000}$ galaxies within the Legacy Survey that are
useful for our analysis.

BOSS is part of the six-year SDSS-III program, that obtained the
spectroscopic redshifts of about $1.5$ million LRGs out to a redshift
of almost $0.7$. We also include data from the eBOSS, which collected
the spectroscopic redshifts of LRGs, Emitting Luminous Red Galaxies
(eLRGs) and quasars (QSO), up to $z = 3.5$. We select galaxies from
these surveys using similar flags\footnote{The flags that end with
\textsc{\textunderscore NOQSO} are specific for the BOSS and eBOSS
galaxies. The description of the flags and their meaning are the same
as in the Legacy Survey.} to the Legacy survey case. We consider
objects which have the flag \textsc{SPECPRIMARY} equal to $1$ and we
select those which have \textsc{CLASS\textunderscore NOQSO} equal to
``Galaxy''. We consider the galaxies with the most reliable redshift
estimations by selecting those which have the flag
\textsc{ZWARNING\textunderscore NOQSO} equal to $0$, $4$ or $16$, and
the flag \textsc{ZERR\textunderscore NOQSO} less than $6 \times
10^{-4}$. Finally, we select the galaxies with redshift between $0.05$
and $0.75$ by using the flag \textsc{Z\textunderscore NOQSO}.  These
selections are applied for the same reasons explained previously for
the Legacy Survey.  We find about $1.9$ million galaxies useful for
our analysis within BOSS and eBOSS.

Table \ref{tab:galselect} shows the selections we made on the galaxy
sample.

\begin{table}[ht]
 \caption{Summary of the considered selections on the galaxy sample.}
    \label{tab:galselect}
    \centering
    \begin{tabular}{|c|c|c|}
        \hline
         Flag & Selection & Justification \\
        \hline
         \textsc{SPECPRIMARY} & 1 & select unique objects \\
         \hline
         \textsc{CLASS} & ``Galaxy'' & select only galaxies \\
         \hline
          \textsc{ZWARNING} & $0$, $4$, $16$ & good spectral fit  \\
         \hline
         \textsc{Z} &  $0.05 < z < 0.75$  & richness-mass relation \\
        \hline
         \textsc{ZERR} &  $<6 \times 10^{-4}$ & accurate redshifts \\
        \hline
    \end{tabular}
   
\end{table}

\begin{figure}
  \centering
  \includegraphics[width=\hsize]{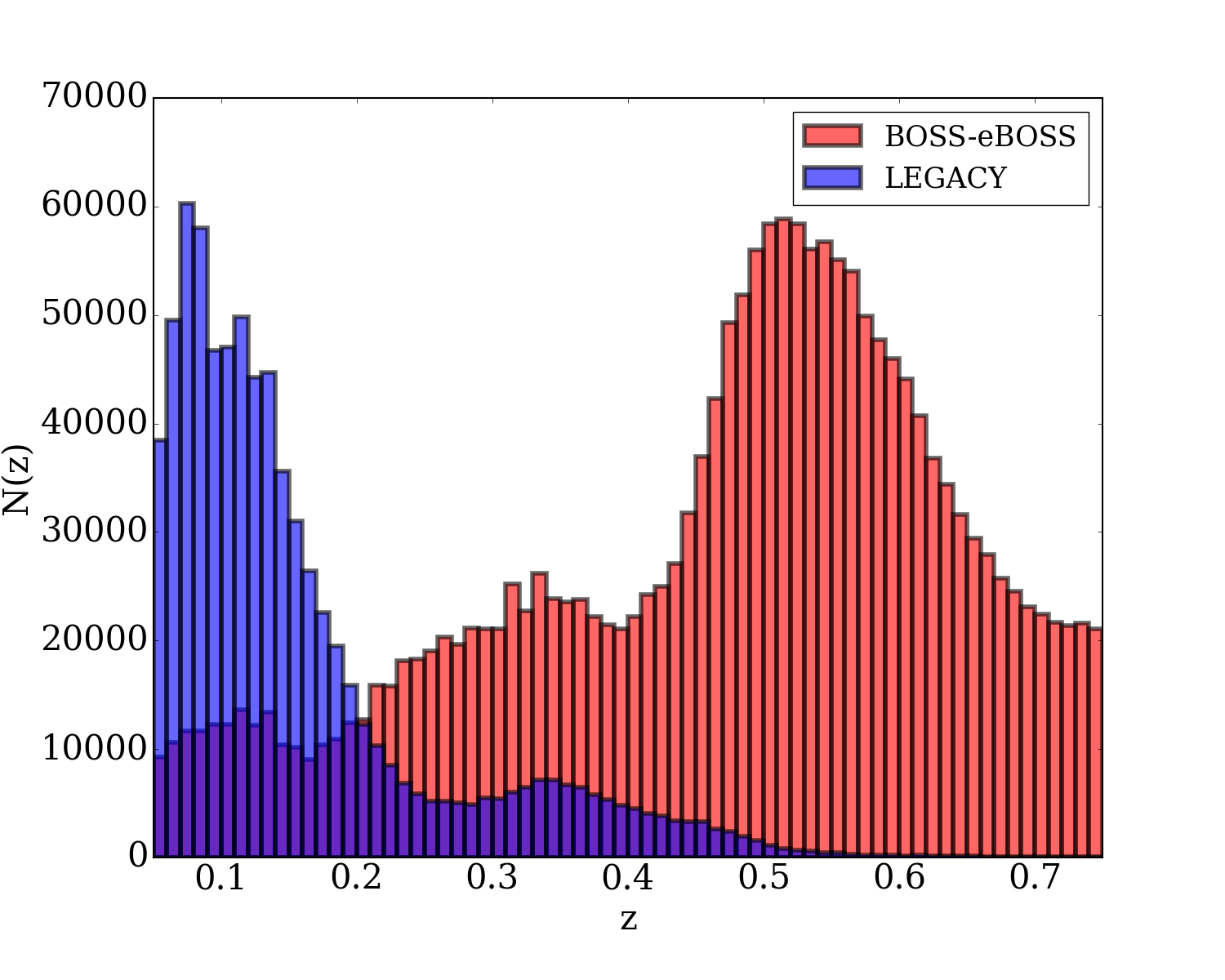}
  \caption{Redshift distribution of the selected galaxies. The blue
    histogram represents the galaxy redshift distribution of the
    Legacy Survey, while the red histogram shows the distribution of
    BOSS and eBOSS.}
  \label{fig:galred}
\end{figure}

Figure \ref{fig:galred} shows the redshift distribution of the
galaxies inside the exploited sample. The mean galaxy redshift within
the Legacy survey is $z \simeq 0.16$, while BOSS and eBOSS galaxies
have a mean redshift of $z \simeq 0.48$.

\section{Searching for cluster member galaxies}\label{searchclustmemb}

To recover the signal of the gravitational redshift effect, it is
necessary to calculate the distribution of the galaxy line-of-sight
velocity offsets, $\Delta$ (see Sec. \ref{model}), as a function of
the distance from the cluster centre. We construct a new catalogue of
cluster member galaxies by cross-correlating the WH15 cluster
catalogue, described in Sec. \ref{whlclustcat}, with the public
spectroscopic galaxy data, described in Sec. \ref{galsample}. Then we
compute the projected transverse distance, $r_{\perp}$, and the
$\Delta$ of all the galaxies with respect to each cluster centre. We
define the latter as the mean value of the angular positions and
redshifts of the BCG closest galaxies, considering objects having a
transverse distance smaller than $r_{500}$ from the BCG. Below we
explain in details how we select our data set.

The WH15 sample provides the BCG angular positions of the identified
clusters. However, most of the BCGs do not have a spectroscopic
redshift measurement. Thus, to increase the number of the available
spectroscopic BCGs, we cross-match the cluster samples with the
considered galaxy catalogue. We take into account the fact that,
according to the SDSS specifications, two galaxies are considered the
same object if they are closer than $3$ arcsec in the Legacy Survey
case, and $2$ arcsec in the BOSS and eBOSS cases. Thus, we consider
the cluster member galaxies only inside clusters which have the BCG
identified in the galaxy sample described in Sec. \ref{galsample}. The
advantage of doing so is that we increase the statistics of the
cluster samples, and we make sure that we only analyse clusters which
have a reliable spectroscopic redshift measurements for their
BCGs. From the cross-matching of the WH15 catalogue and the SDSS data,
we obtain $85\,588$ clusters with a spectroscopic BCG identification;
$47\,779$ of these have the BCG identified in BOSS and eBOSS, while
the other $37\,809$ clusters have the BCG identified in the Legacy
Survey.

Once identified the cluster BCGs, we search for the closest galaxies
to define a new cluster centre. To do this, we compute the projected
transverse distances, $r_{\perp}$, and the line-of-sight velocities of
all the SDSS galaxies with respect to the BCGs. We keep the galaxies
which lie within $r_{\perp} < r_{500}$ and $|\Delta| < 2500$ km
s$^{-1}$ from the BCGs. For each cluster in our sample, we compute the
average value of the redshifts and angular positions of the selected
galaxies, including the BCG, and we define these averages as the new
cluster centres. It should be noted that this centre definition has
never been used in the past literature to measure the gravitational
redshift in galaxy clusters. Instead, it was always assumed that the
cluster centre coincides exactly with the BCG position
\citep{Wojtak_2011, Sadeh_2015, Jimeno_2015}. We find instead that the
average of the member galaxy positions provides a more reliable
location of the centre than the BCG, because the BCG could be
misidentified due to the surface brightness modulation effect. In
fact, peculiar velocities can change the ranking of the two cluster
brightest galaxies. In order to investigate the impact of assuming the
average galaxy positions as the centre of the cluster potential wells,
we compare our results to the ones obtained by assuming instead the
BCG as the cluster centre. We find that the measurements are in
substantial disagreement with the theoretical predictions in the BCG
centre case, showing positive values of the mean of the galaxy
velocity distribution for $r_{\perp}< 2 r_{500}$. Similar results were
obtained by \citet{Jimeno_2015} analysing the WHL12 cluster
sample. The detailed description of the analysis we carried out
assuming the BCG as the cluster centre is presented in Appendix
\ref{BCGresultwhl}.

Once we have defined the cluster centre, we re-select the cluster
member galaxies. We consider a galaxy to be a cluster member if it
lies within a separation of $r_{\perp}<4$ $r_{500}$ and $|\Delta| <
\num{4000}$ km s$^{-1}$ from the cluster centre.  We adopt a lower
limit in the transverse distance, which is about half the size
considered in the past literature works, in order not to depart too
much from the cluster virialized region. On the other hand, this is
the same selection threshold on the galaxy line-of-sight velocity
adopted in \citet{Wojtak_2011} and \citet{Sadeh_2015}.  Hence, we
create a cluster member catalogue, given the galaxy position average
described above as the centre, and we retrieve the galaxy
line-of-sight velocity distribution\footnote{We include the BCG in the
galaxy sample when we calculate the velocity distribution of the
cluster member galaxies.}. It should be noted that the galaxies which
have $|\Delta|$ between $\num{3000}$ km s$^{-1}$ and $\num{4000}$ km
s$^{-1}$ are considered as either foreground or background galaxies,
which are not gravitationally bound to any cluster. Nevertheless, we
include also these galaxies to correct the velocity distribution of
the galaxies which effectively lie within the cluster gravitational
potential well, as described in Sec. \ref{phasespace}.

Before proceeding with the measure of the gravitational redshift, we
make some further selections on the cluster member catalogue. Firstly,
we discard the clusters which have a redshift above $0.5$. We make
this selection in order to avoid the redshift range where the
probability of a false cluster identification is higher than about
$5\%$, as estimated in WHL12 and WH15. Moreover, with this selection
we restrict the analysis to a redshift range where the impact of
possibly incorrect cosmological model parameters is less significant
\citep[see e.g.][]{Wojtak_2011}. Then, we consider only the clusters
which have at least $4$ associated galaxies, and where the average
centre is computed using data of at least $3$ galaxies, including the
BCG. We consider this selection in order to be conservative,
considering only the clusters which have their centres estimated with
a sufficient number of galaxies. When the cluster mass increases, the
gravitational redshift effect becomes stronger and the probability to
have a cluster false identification decreases. Moreover, the galaxy
line-of-sight velocity offsets measured in low-mass clusters are more
affected by the galaxy peculiar velocities than in high-mass clusters
\citep{Kim_2004}. To minimise these possible sources of systematic
uncertainties, we select the clusters which have a mass above $1.5
\times 10^{14}$ M$_{\odot}$. The effects of all these selections are
discussed in Appendix \ref{testselectionwhl}. Finally, we discard the
configurations in which Legacy and BOSS-eBOSS spectra were mixed
together. That is, the cluster member galaxies (comprising the BCGs)
of the Legacy cluster sample are selected only from the Legacy
spectroscopic galaxy sample, while the ones of the BOSS-eBOSS cluster
sample are selected only from the BOSS-eBOSS galaxy sample. We did
this choice because the mixed configurations tend to suppress the
gravitational redshift signal for small values of transverse
distances, as demonstrated by \citet{Sadeh_2015}. It should be noted
that all these conservative selections are possible thanks to the high
statistics of the galaxy and cluster samples we are analysing.
 
Table \ref{tab:clustselect} shows the selections we made on the
cluster sample.
\begin{table}[ht]
 \caption{Summary of the considered selections on the cluster sample.}
    \label{tab:clustselect}
    \centering
    \begin{tabular}{|c|c|c|}
        \hline
             & Selection & Justification \\
        \hline
          redshift & $z<0.5$ & high purity \\
         \hline
         mass & M$_{500}>1.5\times10^{14}$ M$_{\odot}$ & high signal \\
         \hline
         N° of members & $\geq 4$ & high purity \\
         \hline
         N° of members &  & accurate \\
         to compute & $\geq 3$ & cluster centre \\
         the centre & & determination \\
        \hline
        
    \end{tabular}
   
\end{table}  

The final selected sample consists of $\num{3058}$ galaxy clusters and
$\num{49243}$ associated member galaxies. The average redshift is
$\langle z \rangle = 0.25$ and the average mass is $\langle M_{500}
\rangle = 2.75 \times 10^{14}\, \mbox{M}_{\odot}$. The number of
galaxy clusters in our sample is similar to the one of
\citet{Jimeno_2015}. However, we select the cluster member galaxies
using an upper transverse distance limit which is half the size of the
one used in that work. Moreover, to minimise the problems created by
the false cluster identification, we applied more conservative
selections. The average redshift of our sample is similar to those of
the samples analysed by \citet{Wojtak_2011} and \citet {Jimeno_2015},
while we select clusters with higher masses on average.

\begin{figure}
  \centering
  \includegraphics[width=\hsize]{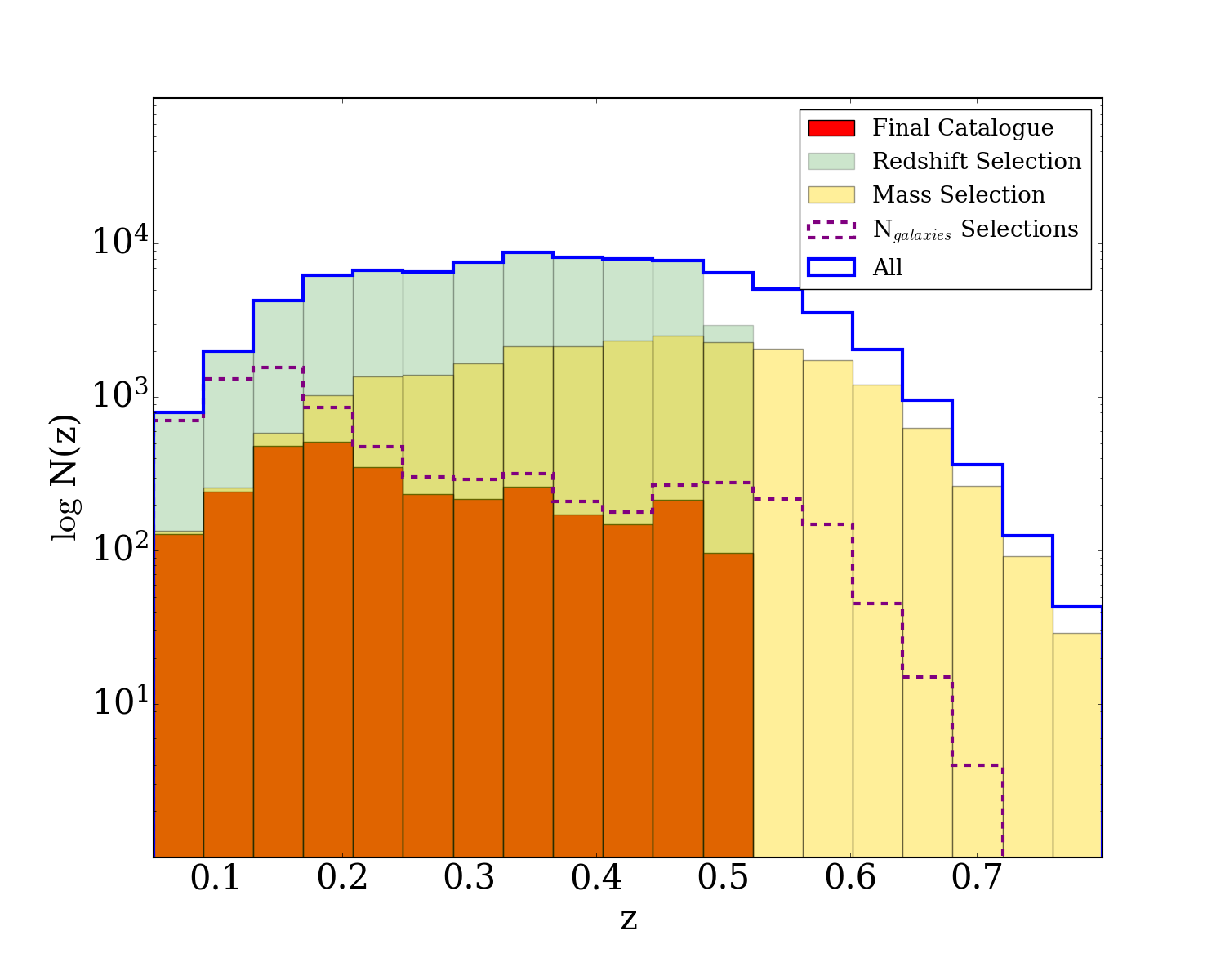}
  \includegraphics[width=\hsize]{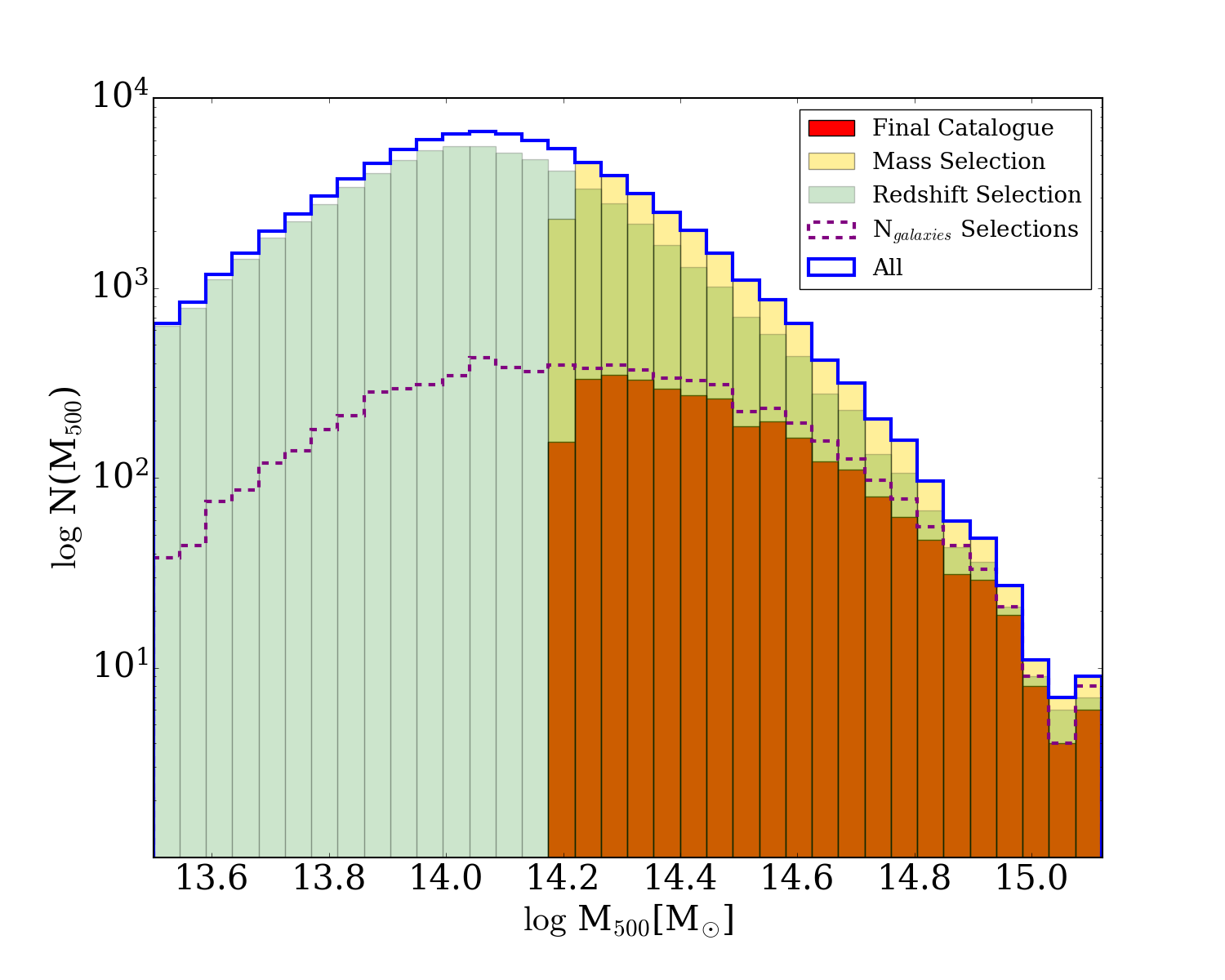}
  \caption{Redshift (top panel) and mass distribution (bottom panel)
    of the selected cluster sample. The blue solid histograms
    represent the distributions of all WL15 clusters within the
    redshift and mass ranges where the richness-mass relation is
    calibrated. The yellow histograms show the distributions of the
    clusters with mass above $1.5 \times 10^{14}$ M$_{\odot}$, while
    the green histograms show the distributions of the clusters with
    $z<0.5$. The dashed purple histograms show the distributions of
    the clusters which have at least 4 associated galaxies. Finally,
    the red histograms represent the distributions of the final
    selected cluster sample.}\label{fig:whlclust}
\end{figure}

Figure \ref{fig:whlclust} shows the redshift and mass distributions of
the WH15 clusters within the redshift and mass ranges where the
richness-mass relation is calibrated, that is $0.05 < z < 0.75$ and $3
\times 10^{13}\,\mbox{M}_{\odot}$. The figure also shows the resulting
distributions after each selection is applied individually, and the
final selected sample analysed in this work.
  
\begin{figure*}
  \centering
  \includegraphics[scale=0.25]{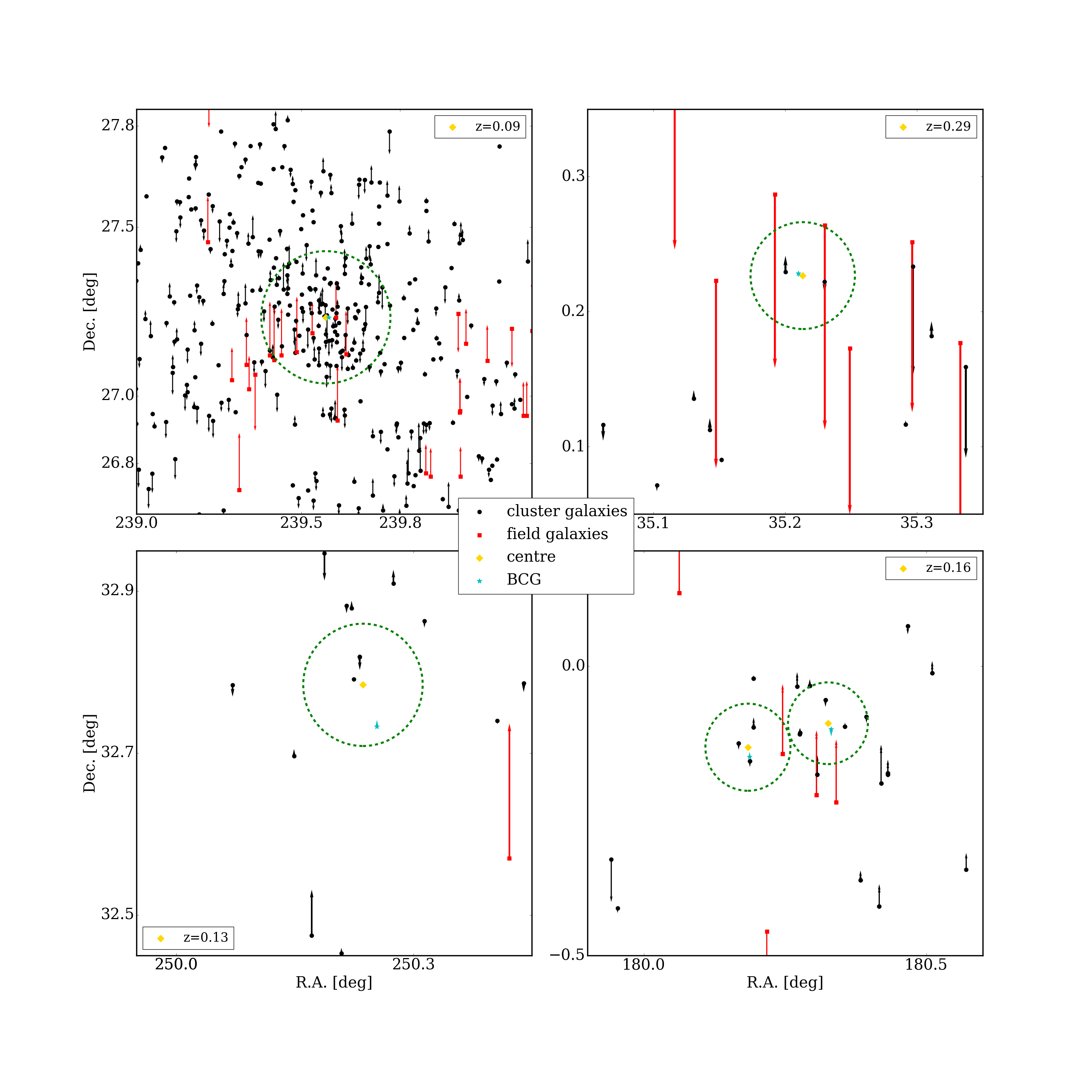}
  \caption{Angular maps around four galaxy clusters of the selected
    sample. The black points represent the cluster member galaxies,
    while the red squares show the positions of the field
    galaxies. The gold diamond shows the cluster centres and the cyan
    star represents the cluster BCGs. The dashed green circle
    indicates the cluster $r_{500}$ radii. The arrows show the galaxy
    line-of-sight velocities with respect to the cluster centres. The
    arrows pointing upward (downward) represent positive (negative)
    velocities. The objects are representative examples of four
    different cluster types: a low redshift massive cluster with a
    large number of identified galaxy members (top left panel); a high
    redshift small cluster, though with a sufficient number of members
    (top right panel); an isolated cluster with only a few identified
    members (bottom left panel); two small close clusters (bottom
    right panel).}
  \label{fig:4clusters}%
\end{figure*}

Figure \ref{fig:4clusters} shows the angular maps around four galaxy
clusters of the final selected sample analysed in this work. Both
member and field galaxies are shown, along with their line-of-sight
velocities. The objects are representative examples of four different
cluster types: a low redshift massive cluster with a large number of
identified galaxy members; a high redshift small cluster though with a
sufficient number of members; an isolated cluster with only a few
identified members; two small close clusters. It should be noted that
the BCG positions are not always near to the cluster centres
identified by the galaxy member positions.

\section{Predicting gravitational redshift in different gravity theories}\label{model}

Gravity theories overall predict that photon frequencies are
redshifted by a gravitational field. When a photon with wavelength
$\lambda$ is emitted inside a gravitational potential $\phi$, it loses
energy when it climbs up in the gravitational potential well, and is
consequently redshifted. The gravitational redshift, $z_g$, observed
at infinity in the weak field limit, can be expressed as follows:
\begin{equation}
  z_g := \frac{\Delta \lambda}{\lambda} \simeq \frac{\Delta
    \phi}{c^2}\,,
\end{equation}
where $\Delta \lambda$ and $\Delta \phi$ are, respectively, the
wavelength and potential differences between the positions where the
photon is emitted and where it is observed.

Let us consider a galaxy, which resides inside a cluster, as a source
of photons. The measurement of the total observed galaxy redshift,
$z_{\rm obs}$, is the sum of different effects, where the main
components are the following: the cosmological redshift, $z_{\rm
  cosm}$, the peculiar redshift, caused by the motion of the galaxy
within the cluster, $z_{\rm pec}$, and the gravitational redshift,
$z_g$:
\begin{equation}
  \ln(1+z_{\rm obs}) =\ln(1+z_{\rm cosm}) + \ln(1+z_{\rm pec}) +
  \ln(1+ z_{\rm g})\,.
\end{equation}
We use differences in the logarithm of the redshifts, as previously
done by \citet{gravfromspider}. \citet{baldry2018} demonstrated that
this provides a better approximation to the galaxy line-of-sight
velocity, with respect to assuming $z=v/c$. The gravitational redshift
depends on the cluster gravitational potential, and thus on the mass
distribution around the galaxy. For a typical cluster mass of
$10^{14}$ $M_{\odot}$, the gravitational redshift is estimated to be
$c z_{\rm g} \simeq 10$ km s$^{-1}$ \citep{cappi,Kim_2004}, which is
about two orders of magnitude smaller than the peculiar redshift. The
tiny effect of the gravitational redshift can be detected only when
the number of analysed galaxies is large enough, that is $N_{\rm gal}
\gtrsim 10^4$ \citep{zhao2013}. Therefore, stacked data of large
samples of clusters and cluster member galaxies are necessary to
measure the gravitational redshift effect with reasonable accuracy.

To disentangle the gravitational redshift from the other components,
we measure the distribution of the galaxy line-of-sight velocities in
the cluster reference frame \citep{Kim_2004}. The line-of-sight
velocity offset is defined as follows \citep{gravfromspider}:
\begin{equation}\label{deltadistr}
  \Delta := c \left[ \ln(1+z_{\rm obs}) - \ln(1+z_{\rm cen})\right]\,,
\end{equation}
where $z_{\rm cen}$ is the redshift of the cluster centre. By
construction, the line-of-sight velocity offset does not depend on the
cosmological redshift component, which is the same in the two terms of
Eq. \eqref{deltadistr}, and thus cancels out. The $\Delta$
distribution of all the galaxy cluster members can be modelled as a
quasi-Gaussian function with nonzero mean velocity,
$\bar{\Delta}$. The value of $\bar{\Delta}$ depends on the spatial
variation of the gravitational potential. This effect is present also
in the most popular alternative theories of gravity, which aim to
modify GR possibly explaining the Universe accelerated expansion
without a dark energy component. Thus, the value of the $\Delta$
distribution mean is the quantity of interest in this
study. Specifically, we will focus on the dependence of $\bar{\Delta}$
on the distance from the cluster centre.

\subsection{General Relativity}\label{grprediction}

The distribution of line-of-sight velocity offsets between cluster
member galaxies and their host cluster centre, defined in
Eq. \eqref{deltadistr}, is expected to have an average value that is
blueshifted \citep{cappi, Kim_2004}. In fact, photons experience the
largest gravitational redshifting at the minimum of the cluster
potential wells, and the gravitational redshift effect decreases
moving towards the cluster outskirts, as the gravitational potential
well decreases as well. Therefore, comparing the redshift of the
cluster centre with the redshifts of member galaxies gives the net
result of a blueshift. For a single galaxy, the gravitational
redshift, expressed as a velocity offset, is given by:
\begin{equation}\label{gravredshiftsinglegal}
  \Delta = \frac{\phi(0)-\phi(r)}{c}\,,
\end{equation}
where $r$ is the distance from the cluster centre. Generally, only the
projected distance from the cluster centre, $r_{\perp}$, is known with
sufficient accuracy. Thus, to compute the gravitational redshift
signal, the density along the line-of-sight to that distance has to be
integrated along with the potential difference.

In this work we assume that the cluster density profile follows the
Navarro-Frank-White radial profile \citep[NFW,][]{NFW}. Moreover, we
use the projected distance from the centre of the cluster in units of
$r_{500}$, because scaling the separation between galaxies and the
associated cluster centres takes advantage of the cluster
self-similarity. Moreover, stacking data by considering comoving
distances is not ideal, as clusters can have a large range of sizes,
and therefore different masses and densities at the same distance from
the centre. The NFW density profile of a cluster, in units of its
radius $r_{500}$, can be expressed as follows \citep{lokasmamon2001}:
\begin{equation}\label{nfwprofiler500}
  \rho(\tilde{r}) = \frac{M_{500} c_{500}^2 g(c_{500})}{4 \pi
    r_{500}^3 \tilde{r}(1+c_{500}\tilde{r})^2}\,,
\end{equation}
where $\tilde{r} := r/r_{500}$, $c_{500}$ is the cluster concentration
parameter defined as $c_{500} := r_{500}/r_s$, $r_s$ is the so-called
\textit{scale radius} of the cluster, and the function $g(c_{500})$
can be expressed as follows:
\begin{equation}\label{g(c)}
  g(c_{500})= \left[ \ln(1+c_{500}) - \frac{c_{500}}{1+c_{500}}
    \right]^{-1}\,.
\end{equation}
\noindent
The gravitational potential, associated with the density distribution
given by Eq. \eqref{nfwprofiler500}, results:
\begin{equation}\label{nfwgravpot}
  \phi(\tilde{r}) = - g(c_{500}) \frac{G M_{500}}{r_{500}}
  \frac{\ln(1+c_{500}\tilde{r})}{\tilde{r}}\,.
\end{equation}
Hence, under these assumptions, the gravitational redshift for a
single cluster (i.e. the mean of the cluster member galaxies velocity
distribution) can be written as follows:
\begin{equation}\label{gravoneclust}
  \bar{\Delta}_{c,gz}(\tilde{r}_{\perp}) = \frac{2 r_{500}}{c
    \Sigma(\tilde{r}_{\perp})} \int_{\tilde{r}_{\perp}}^{\infty}
  \left[\phi(0)-\phi(\tilde{r})\right] \frac{\rho(\tilde{r})\tilde{r}
    \text{d}\tilde{r}}{\sqrt{\tilde{r}^2 - \tilde{r}_{\perp}^2}}\,,
\end{equation}
where $\tilde{r}_{\perp}$ is the projected distance from the centre of
the cluster in units of $r_{500}$. $\Sigma(\tilde{r}_{\perp})$ is the
surface mass density profile computed from the integration of the NFW
density profile along the line-of-sight:
\begin{equation}\label{surfacemassdens}
  \Sigma(\tilde{r}_{\perp})= 2 r_{500}
  \int_{\tilde{r}_{\perp}}^{\infty}
  \frac{\rho(\tilde{r})\tilde{r}}{\sqrt{\tilde{r}^2 -
      \tilde{r}_{\perp}^2}} \text{d}\tilde{r}\,.
\end{equation}
Here we are assuming that a stacked sample of many clusters exhibits
spherical symmetry, even though it is not often the case for a single
cluster \citep{Kim_2004}. Following \citet{Wojtak_2011} the
gravitational redshift signal for a stacked cluster sample can be
calculated by convolving the gravitational redshift profile for a
single cluster with the cluster mass distribution. This operation can
be expressed as follows:
\begin{equation}\label{integralGR}
  \bar{\Delta}_{gz}(\tilde{r}_{\perp}) = \frac{\int_{M_{\rm
        min}}^{M_{\rm max}} \Delta_{c,gz}(\tilde{r}_{\perp})
    \Sigma(\tilde{r}_{\perp})(\text{d}N/\text{d}M_{500})\text{d}M_{500}}{\int_{M_{\rm
        min}}^{M_{\rm max}}
    \Sigma(\tilde{r}_{\perp})(\text{d}N/\text{d}M_{500})\text{d}M_{500}}\,.
\end{equation}
Eq. \eqref{integralGR} can be used to compute the gravitational
redshift effect for a stacked cluster sample as a function of the
projected radius.

Eq. \eqref{integralGR} is valid in any theory of gravity, though
different theories predict different gravitational accelerations
experienced by photons within the clusters. In particular, in
alternative gravity theories the Newtonian constant $G$ is usually
replaced by a function of the cluster radius.

In the following sections we describe the gravitational acceleration
as a function of the cluster radius, $g(r)$, predicted by two
different gravity theories: the $f(R)$ model \citep[see][for a
  complete review]{frtheoryreview} and the Dvali–Gabadadze–Porrati
model \citep[DGP,][]{Dvali2000}. These two alternative gravity
theories appreciably modify the gravity interaction on the largest
scales to reproduce the Universe accelerated expansion, but restore GR
locally, satisfying all current constraints if their parameters are
properly adjusted.

\subsection{The $f(R)$ gravity model}\label{frgravtheory}

In GR, the Einstein-Hilbert action, $S$, which is the integral of the
Lagrangian density over the space-time coordinates, describes the
interaction between matter and gravity and can be expressed as
follows:
\begin{equation}\label{actionLCDM}
  S = \int d^4 x \sqrt{-g} \left[ \frac{M_{pl}^2}{2} (R-2 \Lambda) +
    L_m \right]\,,
\end{equation}
where $M_{pl} = \sqrt{1 / 8 \pi G}$ is the reduced Planck mass, $R$ is
the Ricci scalar, $L_m$ is the matter Lagrangian, and $g$ is the
Friedmann--Lemaître--Robertson--Walker metric determinant.

\citet{STAROBINSKY198099} demonstrates that it is possible to modify
Eq. \eqref{actionLCDM} to describe a consistent gravity theory as
follows:
\begin{equation}
  S = \int d^4 x \sqrt{-g} \left[ \frac{M_{pl}^2}{2} (R-f(R)) + L_m
    \right]\,,
\end{equation}
where the cosmological constant is replaced by a function of the Ricci
scalar, $f(R)$, which is an unknown function. $f(R)$ models are
scalar-tensor theories, where the scalar degree of freedom is given by
$f_R \equiv df/dR$, which mediates the relation between density and
space-time curvature. The theory is stable under perturbations if $f_R
< 0$. The \citet{STAROBINSKY198099} model is constructed to reproduce
the properties of the $\Lambda$CDM framework on linear
scales. Moreover, GR is restored on the smallest scales, thus
fulfilling local constraints.

\citet{schmidt2010} showed that in the strong field scenario,
$|f_{R0}| = 10^{-4}$, the $f(R)$ theory predicts a $4/3$ enhancement
of the gravitational force for all halo masses, that is $G_{f(R)} =
4/3 G$. Thus, the gravitational potential, given by
Eq. \eqref{nfwgravpot}, is significantly enhanced, and the
gravitational redshift effect, given by Eq. \eqref{gravoneclust}, is
consequently stronger than in GR. Following \citet{Wojtak_2011} and
\citet{gravfromspider}, in this work we consider the $f(R)$ theory in
this strong field scenario. Although the strong field scenario
  has been already excluded by different observations
  \citep[e.g.][]{Terukina_2014, Wilcox_2015}, we consider this model
  as comparison because its predictions are significantly enough
  different from GR to be detectable with current gravitational
  redshift measurements.

\subsection{The Dvali–Gabadadze–Porrati gravity model}\label{DGPgravtheory}

In the DGP braneworld scenario \citep{Dvali2000}, matter and radiation
live on a four-dimensional brane embedded in a five-dimensional
Minkowski space. The action is constructed so that on scales larger
than the so-called \textit{crossover scale}, $r_c$, gravity is
five-dimensional, while it becomes four-dimensional on scales smaller
than $r_c$.  Thus, the gravitational potential goes as $1/r$ at short
distances for the sources localised on the brane. As a result, an
observer on the brane will experience Newtonian gravity despite of the
fact that gravity propagates in extra space, which is flat and has an
infinite size. This model admits a homogeneous cosmological solution
on the brane, which obeys to a modified Friedmann equation
\citep{Deffayet_2001}:
\begin{equation}\label{modfried}
  H^2 \pm \frac{H}{r_c} = 8 \pi G \left( \bar{\rho} + \rho_{\rm DE}
  \right)\,,
\end{equation}
where $\rho_{\rm DE}$ is the density associated with the cosmological
constant. The sign on the left-hand side of Eq. \eqref{modfried} is
determined by the choice of the embedding of the brane. The negative
sign is the so-called \textit{self-accelerating} branch, which allows
for accelerated Universe expansion even in the absence of a
cosmological constant. The positive sign is the so-called
\textit{normal} branch, which does not exhibit self-acceleration. On
scales smaller than $r_c$, the DGP models can be described as a
scalar-tensor theory where the brane-bending mode $\varphi$ mediates
an additional attractive (normal branch) or repulsive
(self-accelerating branch) force \citep{Nicolis_2004}.

In DGP models the gravitational forces are governed by the equation:
\begin{equation}
  \nabla \phi = \nabla \phi_N + \frac{1}{2} \nabla \varphi\,,
\end{equation}
where $\nabla \phi_N$ is the Newtonian gravitational potential. It is
possible to find an analytical solution for $\varphi$ in the case of a
spherically symmetric mass. In particular, it is possible to obtain an
equation for the $\varphi$ gradient, which can be expressed as
follows:
\begin{equation}
  \frac{d \varphi}{d r} = \frac{G \delta M(<r)}{r^2} \frac{4}{3 \beta}
  g \left( \frac{r}{r_* (r)} \right)\,,
\end{equation}
where the function $ g \left(y \right)$ is:
\begin{equation}
  g \left( y \right) = y^3 \left[ \sqrt{ 1 + y^{-3}} -
    1 \right]\,,
\end{equation}
and $r_* (r)$ is the so-called \textit{r-dependent Vainsthein
  radius}. The function $r/r_*$ depends on the average overdensity
$\delta \rho (<r)$, within $r$. It is possible to re-scale this
function to a halo with mass $M_{\Delta}$ and radius $R_{\Delta}$,
determined by a fixed overdensity $\Delta$. Thus, we obtain:
\begin{equation}\label{r/rstar}
  \frac{r}{r_* (r)} = (\varepsilon \Delta)^{-1/3} x \left[
  \frac{M(<x)}{M_{\Delta}} \right]^{-1/3}\,,
\end{equation}
where $x := r/R_{\Delta}$ and the quantity $\varepsilon$ is determined by
the background cosmology. By combining these equations,
\citet{schmidt2010} calculated the $g_{\rm DGP}(r)$ parameter, which
quantifies the differences between GR and DGP models:
\begin{equation}\label{gDPGr}
  g_{\rm DGP}(r) = 1 + \frac{2}{3 \beta} g\left( \frac{r}{r_* (r)}
  \right)\,.
\end{equation}
On the largest scales $ g \left( \frac{r}{r_* (r)} \right)$ tends to
$1/2$, so we obtain $g_{\rm DGP} = g_{\rm DGP,lin} = 1 + 1/(3
\beta)$. On the other hand, on the smallest scales where $r \ll r_*$
the modified forces are suppressed.

In this work, we consider a self-accelerating model (sDPG model) with
$\rho_{\rm DE}=0$, and $r_c = 6000$ Mpc, which is adjusted to best
match the constraints derived from Cosmic Microwave Background
observations and Universe expansion history \citep{fang2008}. We made
this choice to test a model which does not need a dark energy
component to explain the Universe accelerated
expansion. \citet{marulli2021} found that the redshift-space
clustering anisotropies of the two-point correlation function of the
same cluster sample exploited in this work are in good agreement with
the predictions of this DGP model.

\citet{schmidt2010} showed that this model predicts a reduction of the
gravitational force, independently of the halo masses. In this work we
set $\beta = - 1.15$ in Eq. \eqref{gDPGr}, and $\varepsilon = 0.32$ in
Eq. \eqref{r/rstar} at $z=0$, in order to reproduce the
\citet{schmidt2010} simulation results.  The model predictions are
significantly affected by the values of the two parameters. In fact,
with $\beta = 1/3$ we recover GR on the largest scales.

\section{Other relativistic and observational effects}\label{GRcorrections}

There are other effects beyond the gravitational redshift that can
cause a shift of the mean of the galaxy velocity distribution, as
shown by \citet{zhao2013} and \citet{kaiser2013}. In this section we
describe all the dominant effects that need to be considered to model
the shift of the mean of the velocity distribution not to bias the
final constraints on the gravity theory. The following description is
valid in any reliable theory of gravity.

\subsection{Transverse Doppler effect}

The peculiar redshift of a galaxy can be decomposed as follows:
\begin{equation}
  1+z_{\rm pec} \simeq 1 + \frac{v_{\rm los}}{c} +
  \frac{1}{2}\frac{v^2}{c^2}\,,
\end{equation}
where $v_{\rm los}$ is the velocity component along the line-of-sight,
and $v$ is the total galaxy velocity. The second-order term, due to
the transverse motion of the galaxy, gives rise to the transverse
Doppler (TD) effect. The TD contributes with a small positive shift of
the mean in the velocity distribution; this is typically of a few
kilometers per second, and is relatively constant with respect to the
distance from the cluster centre. The additional effect on the radial
velocity shift of the mean can be expressed as follows:
\begin{equation}
  \Delta_{\rm TD} = \frac{\langle v_{\rm gal}^2 - v_0^2 \rangle}{2
    c}\,,
\end{equation}
where $v_{\rm gal}$ and $v_0$ are the peculiar velocities of the
galaxies and the cluster centre, respectively. Calculating this effect
involves an integral over the line-of-sight density profile and a
convolution with the mass distribution \citep{zhao2013}. The TD effect
for a single cluster is:
\begin{equation}
  \bar{\Delta}_{\rm c,TD}(\tilde{r}_{\perp}) = \frac{2 Q r_{500}}{c
    \Sigma(\tilde{r}_{\perp})} \int_{\tilde{r}_{\perp}}^{\infty}
  (\tilde{r}^2-\tilde{r}_{\perp}^2)\frac{\text{d}\phi(\tilde{r})}{\text{d}
    \tilde{r}}
  \frac{\rho(\tilde{r})\text{d}\tilde{r}}{\sqrt{\tilde{r}^2 -
      \tilde{r}_{\perp}^2}}\,,
\end{equation}
where $\Sigma(\tilde{r}_{\perp})$ is the surface mass density profile,
given by Eq. \eqref{surfacemassdens}, and $\phi(\tilde{r})$ is the
gravitational potential, given by Eq. \eqref{nfwgravpot}. $Q$ is set
equal to $3/2$ because we assume isotropic galaxy orbits. This
equation must be convolved with the cluster mass function
  of the sample, $\text{d}N/\text{d}M_{500}$, to retrieve the effect
for the stacked cluster sample:
\begin{equation}\label{tdfinal}
  \bar{\Delta}_{\rm TD}(\tilde{r}_{\perp}) = \frac{\int_{M_{\rm
        min}}^{M_{\rm max}} \Delta_{c,TD}(\tilde{r}_{\perp})
    \Sigma(\tilde{r}_{\perp})(\text{d}N/\text{d}M_{500})\text{d}M_{500}}{\int_{M_{\rm
        min}}^{M_{\rm max}}
    \Sigma(\tilde{r}_{\perp})(\text{d}N/\text{d}M_{500})\text{d}M_{500}}\,.
\end{equation}

\subsection{Light-cone effect}

We observe cluster member galaxies which lie in our past light cone
(LC). This causes a bias such that we see more galaxies moving away
from us than moving towards us, as explained by
\citet{kaiser2013}. Hence, this effect causes an asymmetry in the
$\Delta$ distribution, which results in a positive shift of the
mean. The shift caused by the LC effect is:
\begin{equation}
  \Delta_{\rm LC} = \frac{\langle v_{\rm los,gal}^2 - v_{\rm los,0}^2
    \rangle}{c}\,,
\end{equation}
where $v_{\rm los,gal}$ and $v_{\rm los,0}$ are the line-of-sight
velocities of the galaxies and the cluster centre, respectively. The
LC effect is of the same order of the TD effect, and is opposite in
sign relative to the effect of gravitational redshift. To compute the
LC effect on a stacked sample of clusters it is necessary to repeat
the operations already done for the TD effect. Hence, by assuming
isotropic galaxy orbits, we obtain:
\begin{equation}
  \bar{\Delta}_{\rm LC} = \frac{2}{3} \bar{\Delta}_{\rm TD}\,.
\end{equation}

\subsection{Surface brightness modulation effect}

Galaxies in spectroscopic or photometric samples are generally
selected according to their apparent luminosity, $l$. The apparent
luminosity of a galaxy depends on its peculiar motion through the
special relativistic beaming effect, which changes the galaxy surface
brightness (SB) and thus its luminosity. In particular, this effect
enhances the luminosity of galaxies which are in motion towards the
observer, while it decreases the luminosity of those moving
away. Thus, the beaming effect could shift the galaxies moving towards
the observer into the luminosity cut, while it could shift the
galaxies moving away outside the luminosity cut. This causes a bias in
the galaxy selection, promoting galaxies which are moving towards the
observer, with the overall effect of a blueshift on the centre of the
distribution of velocity offsets. Let us consider the effect on the
BCGs. For these galaxies, the flux limit is irrelevant, due to their
high intrinsic luminosity. However, there could be a systematic bias
due to peculiar velocities that can change the ranking of the two
brightest galaxies, possibly causing a wrong selection of the BCG.
This is one of the reasons why we chose not to assume the BCG as the
cluster centre.

The size of the SB modulation effect depends strongly on the galaxy
survey. The relativistic beaming effect can be calculated considering
the fractional change in the apparent galaxy luminosity as a function
of the spectral index, $\alpha$, at the cosmological redshift of the
source, as well as considering the peculiar velocity of the galaxy
\citep{kaiser2013}. The fractional change can be expressed as follows:
\begin{equation}
  \frac{\Delta l}{l} = [3+ \alpha(z)] \frac{v_x}{c}\,.
\end{equation}
\noindent
Furthermore, the modulation of the number density of detectable
objects at a given redshift is given by:
\begin{equation}
  \frac{\Delta l}{l} \delta(z) = - [3+ \alpha(z)] \frac{v_x}{c}
  \frac{\text{d}\ln n (> l_{\rm lim}(z))}{\text{d} \ln l}\,,
\end{equation}
where $\delta(z)$ is the redshift dependent logarithmic derivative of
the number distribution of galaxies and $l_{\rm lim}$ is the apparent
luminosity limit of the survey. The value of $\delta(z)$ depends
strongly on the galaxy sample. The redshift dependence comes from
translating the apparent luminosity limit into an absolute luminosity
limit that varies with redshift. Following \citet{kaiser2013}, we
assume $\alpha(z) = 2$ for the whole redshift range. Hence, assuming
isotropy, we can obtain the predicted shift of the mean due to the SB
effect as follows:
\begin{equation}\label{sbdeltaeff}
  \Delta_{\rm SB} = - 5 \langle \delta(z) \rangle \frac{\langle v_{\rm
      x,gal}^2 - v_{x,0}^2 \rangle }{c}\,,
\end{equation}
where $\langle \delta(z) \rangle$ is the average value of $\delta$
computed over the redshift range of the cluster sample. Just as it was
done for the LC effect, we can write the SB effect as a function of
the TD effect. The result can be expressed as follows:
\begin{equation}\label{sbefffinal}
  \bar{\Delta}_{\rm SB} = - \frac{10}{3} \langle \delta(z) \rangle
  \bar{\Delta}_{\rm TD}\,.
\end{equation}
Thus, we notice that the SB effect is of the same order of the TD and
LC effects, but is opposite in sign. The SB effect leads to a
blueshift of the centre of the distribution of velocity offsets, as
mentioned previously.

\subsection{The combined effect}

The effects described in the previous sections are not the only ones
present, though they are the dominant ones. \citet{cai2017} provided a
comprehensive summary of the different contributions to the mean of
the velocity offset distribution, $\bar{\Delta}$, including the
cross-terms. It is demonstrated that these cross-terms change the
$\bar{\Delta}$ signal by a factor less than $1$ km s$^{-1}$, so they
will not be considered any further in this work. Hence, the
combination of the effects considered in this analysis are the
following:
\begin{equation}\label{deltasum}
  \bar{\Delta} = \bar{\Delta}_{\rm gz} + \bar{\Delta}_{\rm TD} +
  \bar{\Delta}_{\rm LC} + \bar{\Delta}_{\rm SB}\, ,
\end{equation}
which can be written as:
\begin{equation}\label{deltatot}
  \bar{\Delta} = \bar{\Delta}_{\rm gz} + \left(2-5\langle \delta(z)
  \rangle\right) \frac{2}{3} \bar{\Delta}_{\rm TD}\,.
\end{equation}
\noindent
The factor $2/3$ in Eq. \eqref{deltatot} arises from the fact that we
consider logarithmic differences in redshifts, which alters the size
of the TD effect \citep{gravfromspider}.

All the TD, LC and SB effects are small compared to the
  gravitational one. Thus, for simplicity, as previously done in past
  literature works, we will refer to the combined effect as the
  gravitational redshift effect.

\section{Computing the theoretical models}\label{theoreticalmodels}

We use Eq. \eqref{deltatot} to predict the mean value of the member
galaxy velocity distribution in the different theories of gravity
considered in this work. Specifically, we calculate $\bar{\Delta}_{\rm
  gz}$ and $\bar{\Delta}_{\rm TD}$, given by Eq. \eqref{deltatot}, as
well as $\langle \delta(z) \rangle$, given by
Eq. \eqref{sbefffinal}. We compute $\bar{\Delta}_{\rm gz}$ by solving
Eq. \eqref{integralGR}, while $\bar{\Delta}_{\rm TD}$ is computed with
Eq. \eqref{tdfinal}. The red histogram in the bottom panel of
  Fig. \ref{fig:whlclust} shows the measured cluster mass distribution
  used to compute the integrals in Eqs. \eqref{integralGR} and
  \eqref{tdfinal}, where the minimum and maximum masses of the samples
  are $1.5 \times 10^{14}$ M$_{\odot}$ and $2 \times 10^{15}$
  M$_{\odot}$, respectively. The \citet{Duffy2008} relation and the
  NFW density profile are used to compute the $c_{500}$ concentration
  parameter for each cluster. The median value of $c_{500}$ of the
  selected cluster sample is about $2.5$, which is agreement with the
  typical value expected for clusters in these ranges of mass and
  redshift \citep{Miyazaki_2017}. 

We follow the procedure described in \citet{kaiser2013} to compute the
intensity of the galaxy number distribution. Specifically, we
calculate the redshift-dependent logarithmic derivative of the number
distribution of galaxies, $\delta(z)$, defined as:
\begin{equation}
  \delta(z) := \frac{\text{d} \log n (< M_{\rm lim}(z))}{\text{d} \log
    M},
\end{equation}
where $M_{\rm lim}$ is the absolute magnitude limit of the galaxy
survey.

Following \citet{kaiser2013} and \citet{Jimeno_2015}, we use the model
of \citet{monterolumfunc} for the \textit{r}-band luminosity function,
that is a Schechter function with a characteristic magnitude $M_* -
5\log_{10}h = -20.7$ and a faint end slope $\alpha = -1.26$.  Ideally,
we should use a specific luminosity function of the cluster member
galaxies. However, \citet{Hansen_2009} demonstrated that the
parameters of the overall luminosity function for the cluster member
galaxies does not differ significantly from the ones of the luminosity
function of the field galaxies. Thus, we assume an overall luminosity
function, as it was done in the past literature
\citep[][]{kaiser2013,Jimeno_2015,gravfromspider}.

To calculate $\delta(z)$, we use the SDSS fibre magnitude limit in
\textit{r-band} of $22.29$ \footnote{ This value is taken from the
  SDSS official web site, see
  https://www.sdss.org/dr12/algorithms/magnitudes}, as magnitude
cut. The intensity of the SB effect is computed from the average value
of $\delta(z)$ over the cluster sample redshift range by solving the
following integral:
\begin{equation}\label{integraldzmedio}
  \langle \delta(z) \rangle = \frac{\int_{z_1}^{z_2} \delta(z)
    (\text{d}N / \text{d}z) \text{d}z}{\int_{z_1}^{z_2} (\text{d}N /
    \text{d}z) \text{d}z}\,,
\end{equation}
where $z_1=0.05$ and $z_2=0.5$ are the lowest and upper redshift
limits of the exploited cluster samples. To solve the integral in
Eq. \eqref{integraldzmedio} we consider, as $\text{d}N/\text{d}z$, the
galaxy redshift distribution shown in Fig. \ref{fig:galred}. The result
of the integral computation is: $\langle \delta(z) \rangle = 0.516$.

\begin{figure}
  \centering
  \includegraphics[width=\hsize]{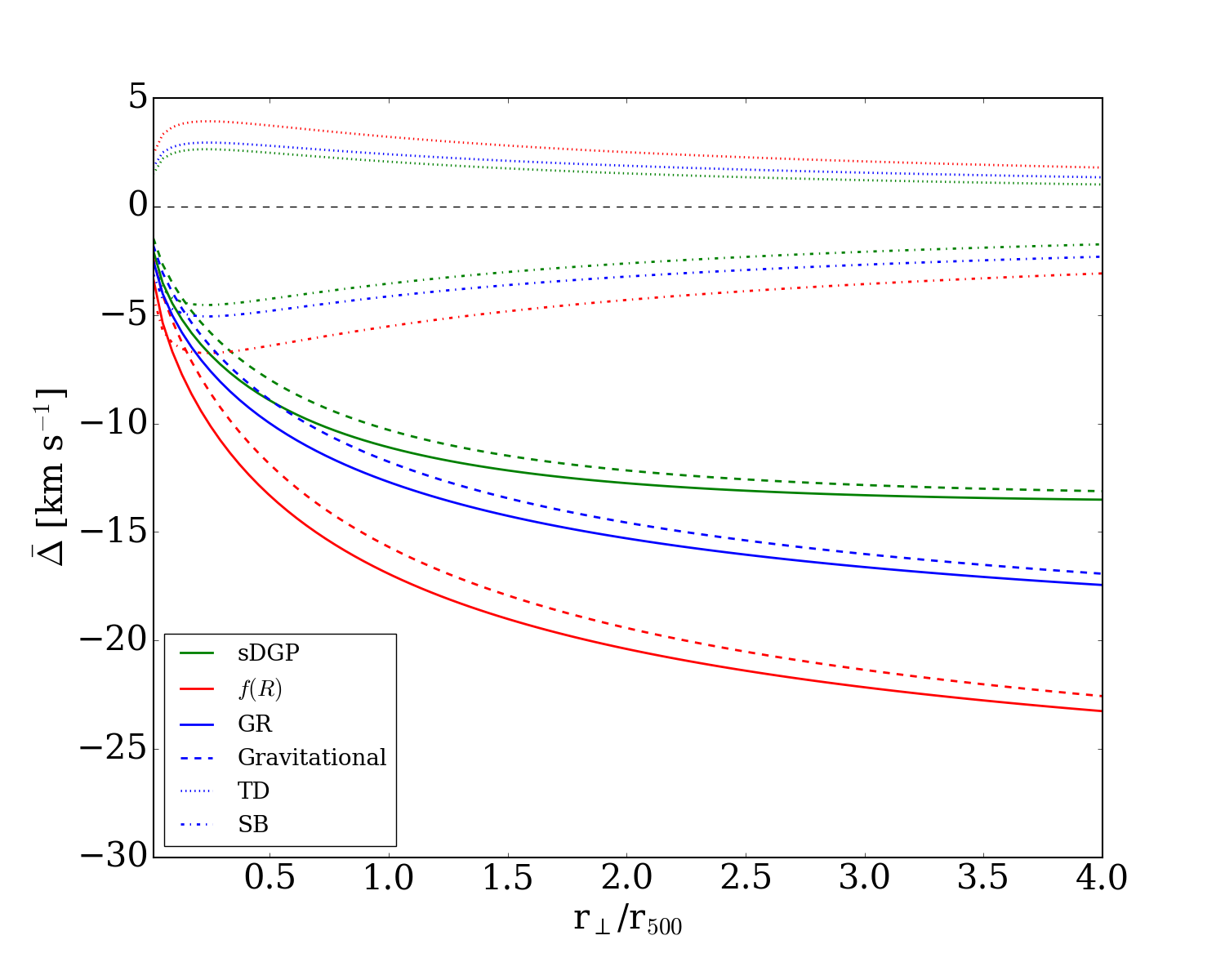}
  \caption{Predicted value of $\bar{\Delta}$ as a function of the
    cluster radius in units of $r_{500}$. The blue lines refer to the
    predictions computed assuming GR, the green lines refer to the
    sDGP predictions and the red lines refer to $f(R)$ gravity
    theory. For each colour, the dashed line shows the gravitational
    effect only, the dotted line the TD effect, the dot-dashed line
    the SB effect, and the solid line is the combined effect.}
  \label{fig:deltapredctirad}%
\end{figure}

Figure \ref{fig:deltapredctirad} shows the predicted value of
$\bar{\Delta}$ as a function of the transverse distance from the
cluster centre in units of $r_{500}$. The figure shows not only the
combined effect given by Eq. \eqref{deltatot} (solid lines), but also
the gravitational, TD and SB effects individually. The $\bar{\Delta}$
value becomes more negative as the transverse distance increases. This
is expected because $\bar{\Delta}$ gives information on the difference
between the gravitational potential at the cluster centre and at a
given transverse distance from it. This difference increases going
outside the cluster potential well, then the shift of the mean of the
cluster member galaxy velocity distribution grows. Figure
\ref{fig:deltapredctirad} also shows that the TD effect causes a
positive shift of $\bar{\Delta}$, while the SB effect causes a
negative shift, as described in Sec. \ref{GRcorrections}. The TD and
SB effects are small compared to the gravitational effect, as
expected. Particularly, in GR at $4 r_{500}$ from the cluster centre
the TD and SB effects have an intensity of about $2$ km s$^{-1}$ and
$-2.5$ km s$^{-1}$ respectively, while the gravitational effect has a
magnitude of about $-15$ km s$^{-1}$.  Furthermore, as the distance
from the cluster centre increases, the difference between GR, $f(R)$
and sDGP predictions rises as well.

\section{Measuring the gravitational redshift}\label{result}

\subsection{Correction of the phase-space diagram}\label{phasespace}
\begin{figure}
  \centering
  \includegraphics[width=\hsize]{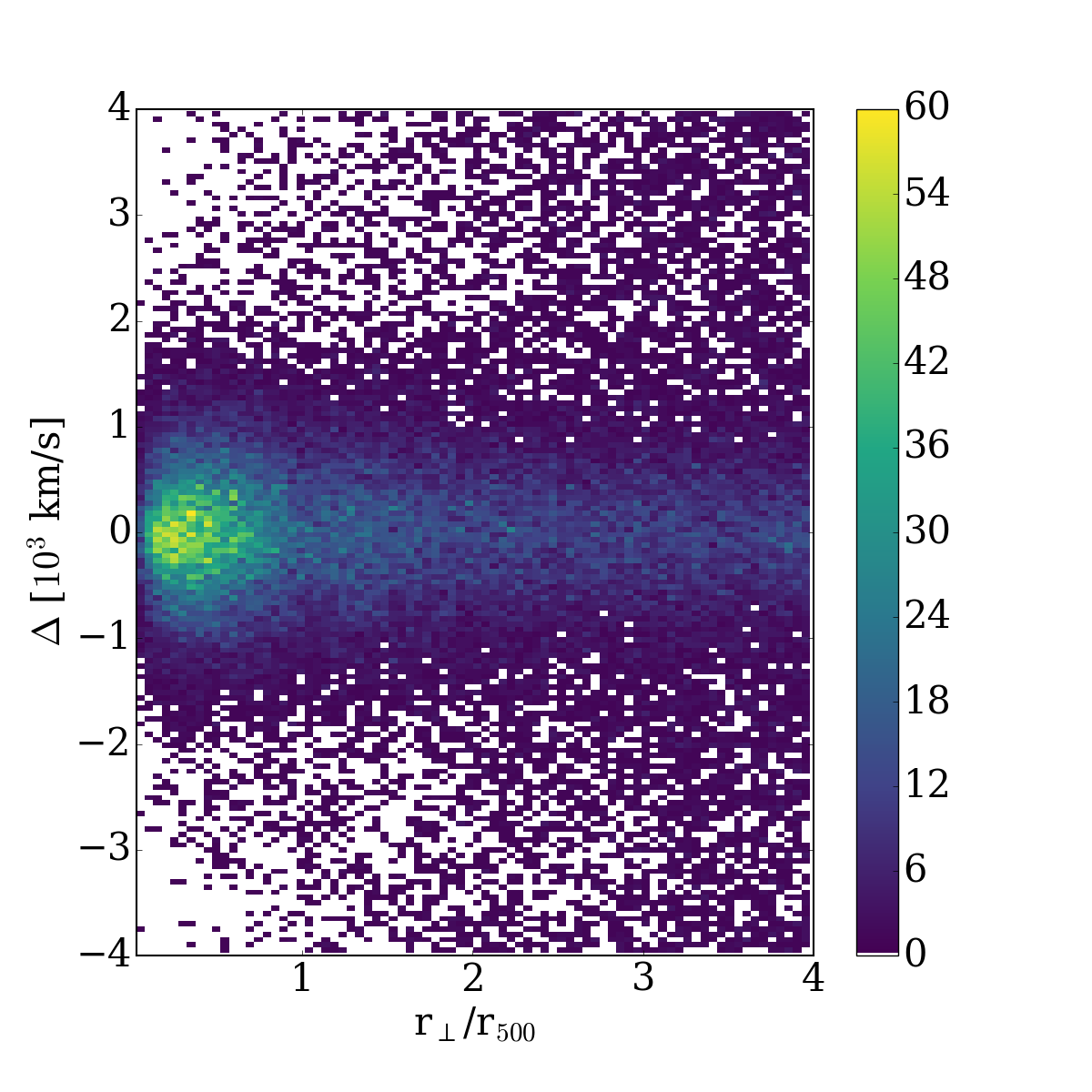}
  \includegraphics[width=\hsize]{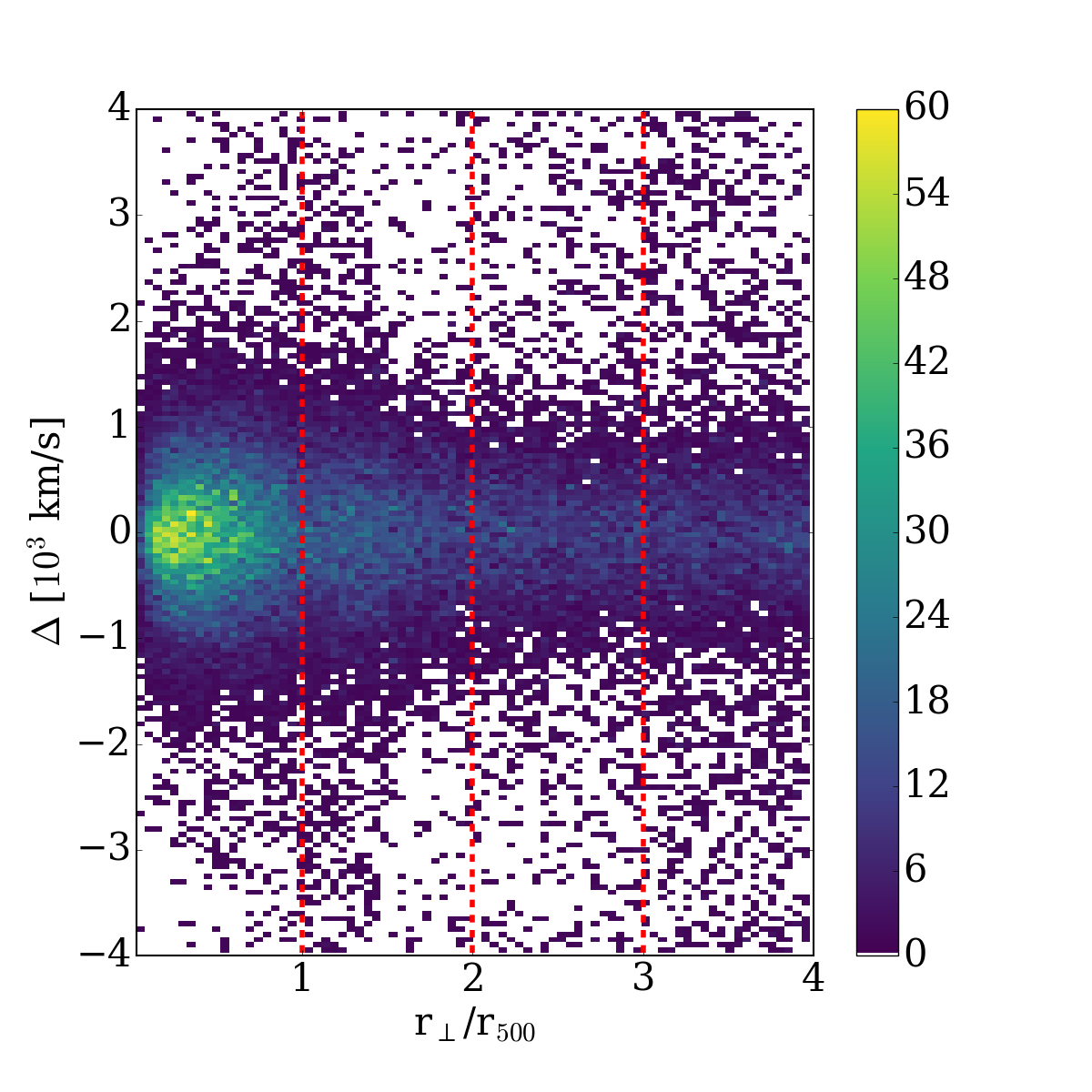}
  \caption{Phase-space diagram for the stacked member galaxy data of
    the clusters in the WH15 catalogue. The
    top panel shows the diagram before the correction procedure, while
    the bottom panel shows the background-corrected phase-space
    diagram. The colour bar shows the number of member galaxies we
    have in each bin. The bins have a size of $0.05$ $r_{500}$
    $\times$ $50$ km s$^{-1}$. The vertical red dashed lines show the
    bins where we calculate the mean of the velocity distributions.}
  \label{fig:whlphasespace}%
\end{figure}

To measure the gravitational redshift effect from the cluster member
catalogue constructed in Sec. \ref{searchclustmemb}, we stack all the
data of the member galaxies (i.e. the transverse distances
$r_{\perp}$, and line-of-sight velocities $\Delta$) in a single
phase-space diagram \citep{Kim_2004,Wojtak_2011}. Figure
\ref{fig:whlphasespace} (top panel) shows the stacked line-of-sight
velocity offset distributions for the member galaxies in the WH15
catalogue.

The phase-space diagram is affected by the contamination of the
foreground and background galaxies, which are not gravitationally
bound to any selected cluster, due to projection effects. We have to
take into account only the galaxies that are within the cluster
gravitational potential well to make a reliable measurement of the
gravitational redshift. We follow the procedure described in
\citet{Jimeno_2015} to remove the contamination of foreground and
background galaxies. The galaxies which do not belong to any cluster
are considered statistically, once the data of all cluster member
galaxies have been stacked into a single phase-space diagram
\citep[see][for a detailed review on foreground and background galaxy
  removal techniques]{wojtak2007}. Firstly, we split the phase-space
distribution in bins of size $0.05$ $r_{500}$ $\times$ $50$ km
s$^{-1}$. We assume that the galaxies which lie in the stripes
$\num{3000}$ km s$^{-1}$ $< |\Delta| <$ $\num{4000}$ km s$^{-1}$
belong either to pure foreground or to pure background, as already
described in Sec. \ref{searchclustmemb}. We select the upper limit for
galaxy line-of sight velocity of $\num{4000}$ km s$^{-1}$ following
the past literature work of \citet{Wojtak_2011} and
\citet{Sadeh_2015}. On the other hand, we made some tests changing the
lower limit of $\num{3000}$ km s$^{-1}$.  Selecting the lower cut-off
within the range $\num{2000}$ km s$^{-1}$ $< |\Delta| <$ $\num{3500}$
km s$^{-1}$, we obtain consistent results, within the errors, to the
ones we present in Sec. \ref{measures}.

We fit a quadratic polynomial function, which depends on both $\Delta$
and $r_{\perp}$, to the points in both stripes. We use the
interpolated model to correct the phase-space region where $|\Delta|$
is less than $\num{3000}$ km s$^{-1}$, namely the region where the
galaxies are gravitationally bound. The function
$f(r_{\perp},\Delta)$, that we use to model the phase-space region
where the background and foreground galaxies lie, can be expressed as
follows:
\begin{equation}
  f(r_{\perp},\Delta) = a r_{\perp}^2 + b \Delta^2 + c 
  r_{\perp} \Delta + d \Delta + e r_{\perp} + f\,,
\end{equation}
where the $a$, $b$, $c$, $d$, $e$, $f$ represent the free
parameters of the model.  We use a function that depends on both
$r_{\perp}$ and $\Delta$ because, due to observational selections, we
may observe more galaxies which are close to us with respect to the
cluster centre (i.e. negative $\Delta$) than further away
(i.e. positive $\Delta$). Moreover, the possibility to find galaxies
that do not belong to the cluster increases with the distance from the
cluster centre.

The bottom panel of Fig. \ref{fig:whlphasespace} shows the
background-corrected phase-space diagram. After removing the
background, the phase-space diagram clearly shows the inner region
where the gravitationally bound galaxies reside. Indeed, most of the
galaxies in the foreground and background regions have been removed,
proving that the background-correction method succeeded. However, not
all the contamination has been removed because of the intrinsic
uncertainties in the fitting. Thus, we take into account this error
when we fit the galaxy velocity distributions. In particular, we
consider the mean \textit{rms} as the error of the fitting procedure. 
In fact, the bottom panel of Fig. \ref{fig:whlphasespace}
  shows that a certain amount of galaxies with a large velocity offset
  around $r_{\perp}/r_{500} \sim 1$ and $3<r_{\perp}/r_{500}<4$ is
  still present after the background correction. Nevertheless, in each
  bin of these parameter regions we find at most one galaxy. This non-uniform 
  background subtraction might be due to minor statistical
  uncertainties. To test the impact of this effect on the final
  results of our analysis, we measured again the gravitational
  redshift considering only those galaxies with $|\Delta| <
  \num{2000}$ km s$^{-1}$, finding consistent results, within the
  errors, to the ones presented in Sec. \ref{measures}.

\subsection{Fitting the data}\label{fit}

We split the background-corrected phase-space diagrams into four equal
bins of transverse distance to recover the gravitational redshift
signal as a function of the transverse distance from the cluster
centre. Each bin has a width of $1 r_{500}$, as shown in
Fig. \ref{fig:whlphasespace}. We fit the galaxy line-of-sight velocity
distribution within each bin, in order to recover the mean of the
distribution, $\bar{\Delta}$. The mean value of the distribution is
proportional to the intensity of the gravitational redshift effect and
we expect a negative value, as explained in Sec. \ref{model}. We
perform a Monte Carlo Markov Chain (MCMC) statistical analysis to fit
$\bar{\Delta}$ within each bin. We model the velocity distribution as
a double Gaussian function, which can be expressed as follows:
\begin{equation}\label{modelfdelta}
    f(\Delta) = \frac{\varepsilon}{\sqrt{2 \pi \sigma_1^2}} \exp
    \left[ {\frac{(\Delta-\bar{\Delta})^2}{2 \sigma_1^2}} \right] +
    \frac{1 - \varepsilon}{\sqrt{2 \pi \sigma_2^2}} \exp \left[
      {\frac{(\Delta-\bar{\Delta})^2}{2 \sigma_2^2}} \right]\,,
\end{equation}
where the two Gaussian functions have the same mean,
$\bar{\Delta}$. The relative normalisation of the two functions,
$\varepsilon$, and their widths, $\sigma_1$ and $\sigma_2$, are
considered as free parameters of the MCMC analysis, and marginalised
over. The Bayesan fit is implemented by using a Gaussian likelihood,
with flat priors on all the free model parameters. We consider the
combination of two independent sources of errors: i) the Poisson noise
and ii) the error of the background-correction method. The
quasi-Gaussian function given by Eq. \eqref{modelfdelta} takes into
account for the intrinsic non-Gaussian distributions of galaxy
velocities in individual clusters and for the different cluster
masses.

\section{Results}\label{measures}

\begin{figure}
   \centering
   \includegraphics[width=\hsize]{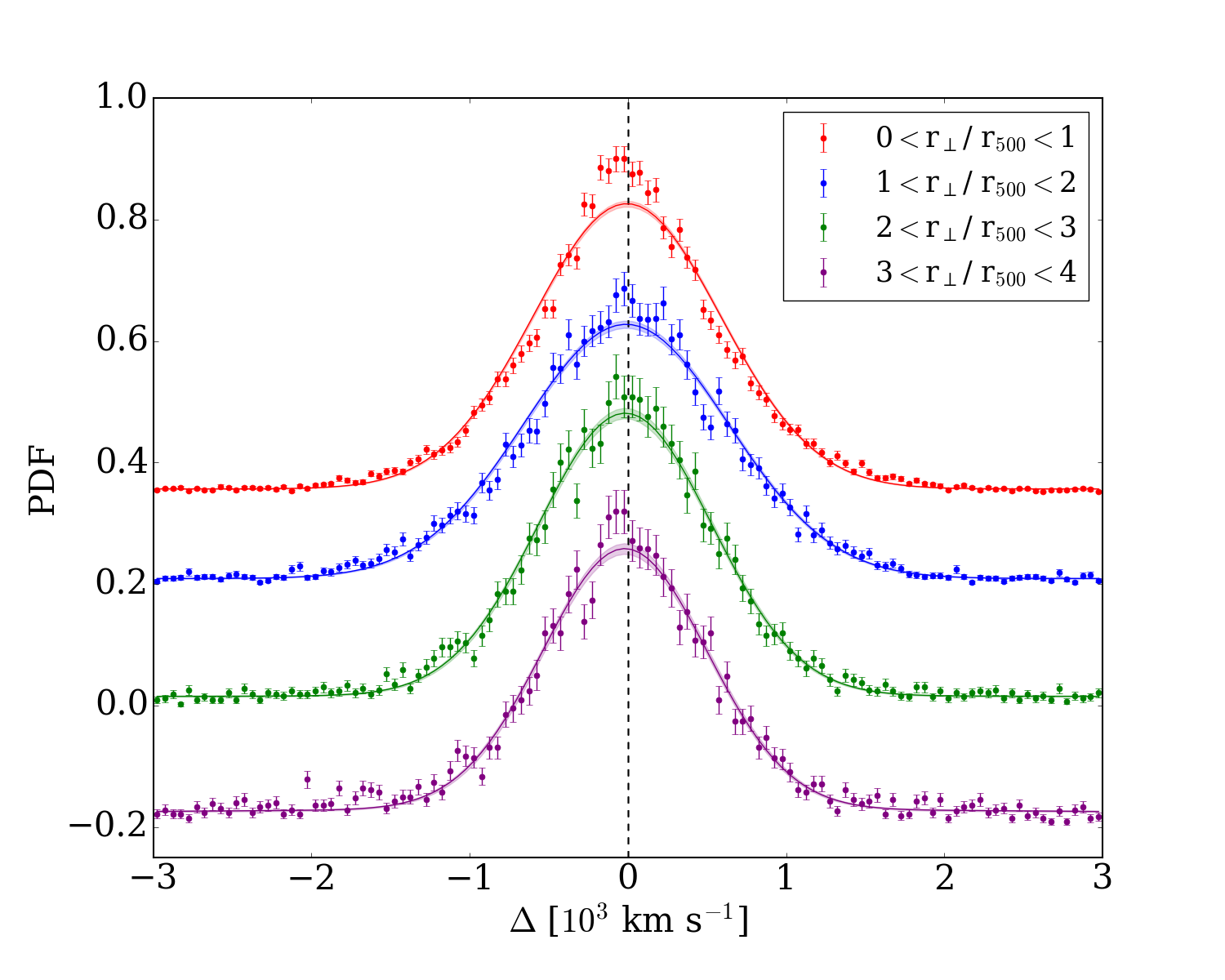}
   \caption{Velocity distributions of the WH15 cluster member galaxies
     in the four bins of projected transverse distance. These
     distributions are shifted vertically by an arbitrary amount
     ($-0.2$, $0$, $0.2$ and $0.35$), for visual purposes. The
     coloured points represent the data of the binned
     background-corrected phase-space diagram and the error bars
     represent the Poisson noise combined with the error of the
     background-correction method. The solid lines and the shaded
     coloured areas show the best-fit models, and their errors,
     respectively.}
    \label{fig:diswhln>2mean}
    \end{figure}

Figure \ref{fig:diswhln>2mean} shows the velocity distributions in
each bin of projected transverse distance. The figure shows the data
of the binned background-corrected phase-space diagram, the associated
error bars, and the best-fit models within each bin. We notice that
the model systematically underestimates the data with low $\Delta$ at
any distance from the centre. This is a feature that was present also
in the past literature works \citep{Wojtak_2011, Jimeno_2015}, and
does not significantly affect the final $\bar{\Delta}$ estimation.

\begin{figure}
   \centering
    \includegraphics[width=\hsize]{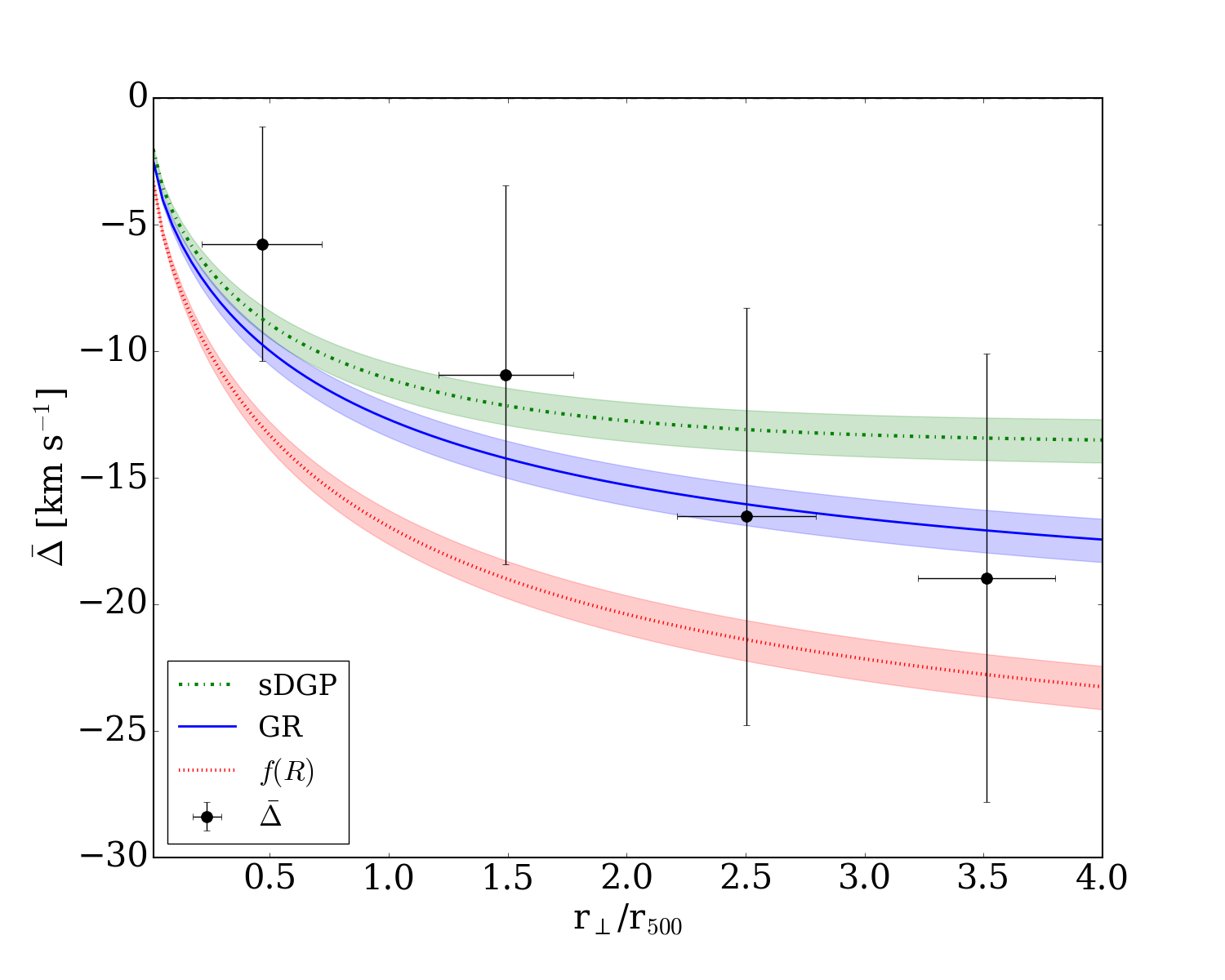}
    \caption{Comparison between the estimated $\bar{\Delta}$ of the
      WH15 cluster member galaxies within each bin of transverse
      distance and the theoretical predictions from GR (blue solid
      line), $f(R)$ (red dotted line), sDGP (green dash-dotted
      line). The shaded coloured areas show the model errors which are
      caused by the fitting uncertainties on the cluster mass
      distribution, and the dispersion of the cluster redshifts. The
      black points show the estimated $\bar{\Delta}$. The vertical
      error bars represent the range of $\bar{\Delta}$ parameter
      containing $68 \%$ of the marginalised posterior probability,
      while the horizontal error bars show the dispersion of the
      galaxy transverse distances in a given bin.}
    \label{fig:whln>2misuram>14.2}
    \end{figure}

Figure \ref{fig:whln>2misuram>14.2} shows the comparison between the
estimated $\bar{\Delta}$ within each bin and the GR, $f(R)$ and sDGP
theoretical predictions, as a function of the transverse distance from
the cluster centre. As it can be seen, we find a clear negative shift
of the mean of the velocity distributions, as we expected from the
theoretical analysis described in Sec. \ref{model}. As shown in
Fig. \ref{fig:whln>2misuram>14.2}, our measurements are in agreement,
within the errors, with the predictions of GR and sDGP, while in
marginal tension with $f(R)$ predictions, though the disagreement is
not statistically significant.

The richness-mass scaling relation is a crucial element in this kind
of analyses, since it can introduce systematic biases in the final
constraints if not properly calibrated in the assumed gravity theory
considered. In particular, the so-called fifth force possibily arising
from the new scalar degrees of freedom of modified gravity models,
such as the $f(R)$ scenario considered in this work, modifies the
relation between the cluster masses and observable proxies
\citep{Terukina_2014, Wilcox_2015}. To test the impact of this effect,
we performed the full analysis again using the Eqs. (6) and (22) from
\citet{Mitchell_2021} to compute the M$_{500}$ masses for each cluster
in the $f(R)$ strong field scenario. We find that the new M$_{500}$
values are on average about $15\%$ higher than the corresponding
masses estimated in GR, and the new gravitational redshift
measurements are shifted on averaged by about $20\%$ towards positive
$\bar{\Delta}$ values with respect to the results shown in
Fig. \ref{fig:whln>2misuram>14.2}. The results of this test
  are shown in Appendix \ref{FRresult}. These new results are within
the estimated statistical uncertainties, thus not introducing dominant
systematic effects for the current analysis. An accurate calibration
of the mass-observable scaling relation in different modified gravity
models will be mandatory for similar analyses on next-generation large
cluster samples and will deserve a detailed study which is beyond the
scope of the present paper.

We measure also the integrated gravitational redshift signal up to $4
r_{500}$, $\bar{\Delta}_{int}$, by considering all the cluster member
galaxies in the background-corrected phase-space diagram shown in
Fig. \ref{fig:whlphasespace}. We obtain $\bar{\Delta}_{int} = -11.4
\pm 3.3$ km s$^{-1}$, which is in agreement, within the errors, with
the expected value of $-10$ km s$^{-1}$ predicted in GR for clusters
in the mass range of the WH15 cluster member catalogue.

We fit the measured value of $\bar{\Delta}$, shown in
Fig. \ref{fig:whln>2misuram>14.2}, to impose new constraints on the
gravity theory and discriminate among the three different models
considered. To do this, we modify the theoretical model given by
Eq. \eqref{deltatot}, by changing the gravitational acceleration
experienced by the photons inside the clusters. In practice, we
multiply the gravitational constant $G$ by a constant $\alpha$, which
will be considered as the free parameter of the fit. This simple model
is accurate enough to take into account the modification of the
gravitational force predicted by both $f(R)$ and sDGP models. By
construction, $\alpha$ is equal to unity in GR theory, while $\alpha =
1.33$ in the $f(R)$ theory and $\alpha \simeq 0.85$ in the sDGP
model. We perform a MCMC analysis to fit the measured $\bar{\Delta}$,
using a Gaussian likelihood. It should be noted that this fitting
procedure has never been implemented in past literature works.

\begin{figure}
   \centering
   \includegraphics[width=\hsize]{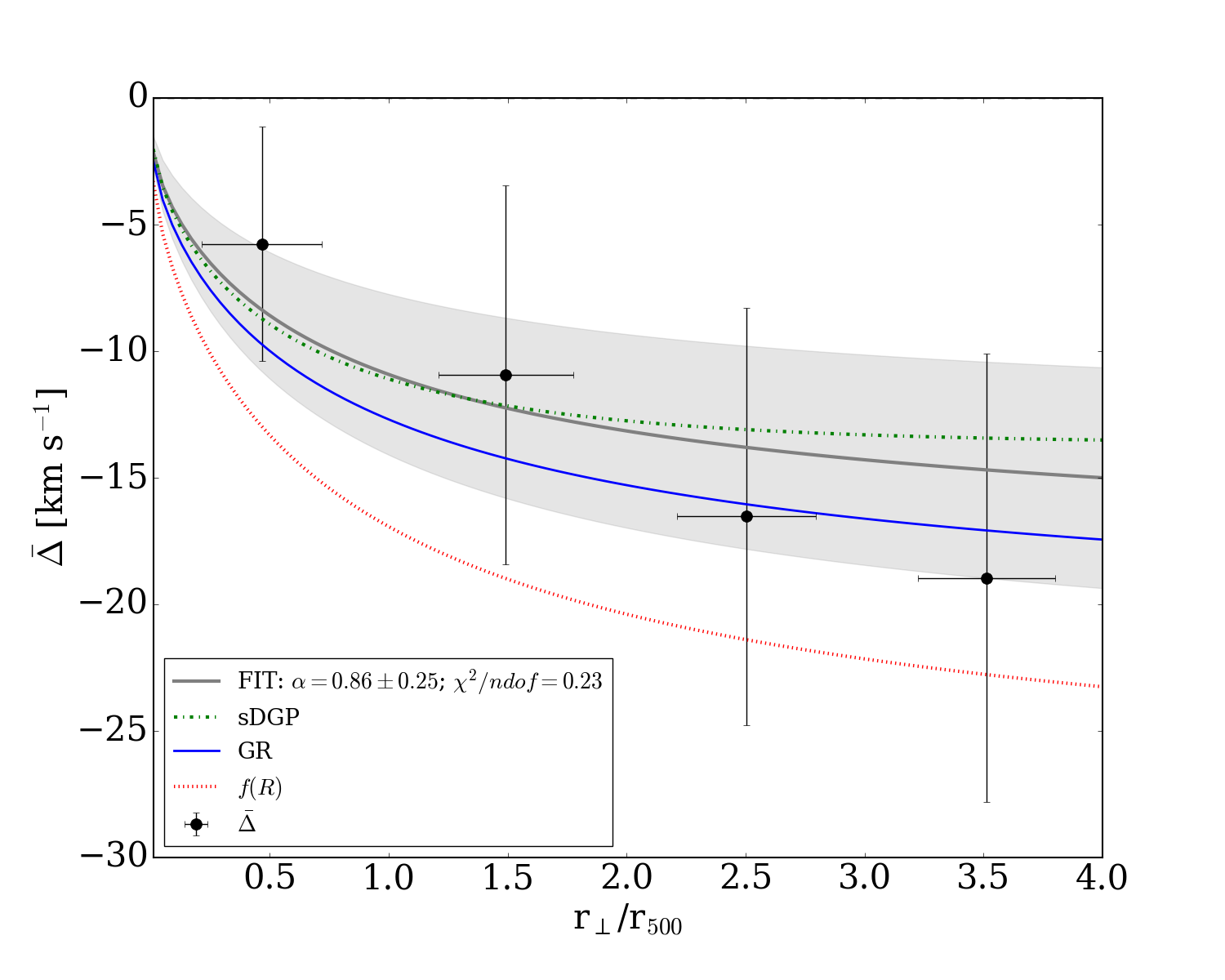}
    \caption{Best-fit model of $\bar{\Delta}(r_{\perp}/r_{500})$ from
      MCMC (grey solid line) for the WH15 cluster member galaxies. The
      shaded grey area shows the $68 \%$ uncertainty on the posterior
      median. The theoretical predictions of GR (blue solid line),
      $f(R)$ (red dotted line), sDGP (green dash-dotted line) are
      shown for comparison.}
    \label{fig:whln>2fitm>14.2}
    \end{figure}

Figure \ref{fig:whln>2fitm>14.2} shows the results of the MCMC
analysis. We obtain $\alpha = 0.86 \pm 0.25$, with a reduced $\chi^2 =
0.23$. The value of the reduced $\chi^2$ indicates a possible
overestimation of the measurement errors. The best-fit results confirm
that our measurements are in agreement, within the error, with GR and
sDGP predictions, while the $f(R)$ model is marginally discarded at
about $2 \sigma$. This result is consistent with past
  literature works that have already discarded the $f(R)$ strong field
  scenario considered here \citep[e.g.][]{Terukina_2014,
    Wilcox_2015}. Our result is also compliant with what has been
found by \citet{marulli2021} from a redshift-space distortion analysis
of the two-point correlation function of the same cluster catalogue.
On the other hand this is in slight disagreement with the
past literature works of \citet{Wojtak_2011} and
\citet{gravfromspider}, whose results were consistent also with
$f(R)$. Nevertheless, as noted above, a proper self-consistent
  treatment of the richness-mass scaling relation in modified gravity
  models would be required to assess unbiased constraints. In fact,
  the analysis presented here provides robust constraints on GR
  predictions, for which the likelihood model calibration is accurate
  enough, given the current uncertainties. On the other hand, the
  comparison of our measurements with different gravity theories
  should be taken with caution, and no strong conclusions can be drawn
  in this respect.

\section{Conclusions}\label{conclusions}

In this work we tested the Einstein theory of GR by measuring the
gravitational redshift effect in galaxy clusters, within the
$\Lambda$CDM cosmological framework. To perform the gravitational
redshift measurements, we constructed a new cluster member galaxy
catalogue, as discussed in Sec. \ref{searchclustmemb}. Differently
from the past literature works, we used the average positions and
redshifts of central galaxies to estimate the cluster centres.  In
Appendix \ref{BCGresultwhl} we compare the results obtained with this
choice to those obtained assuming the BCGs as the cluster
centres. Following the method described by \citet{Kim_2004}, we
stacked the data of the cluster member galaxies in a single
phase-space diagram and corrected it for the background and foreground
galaxy contaminations, as explained in Sec. \ref{phasespace}. We
splitted the phase-space diagrams in four bins of transverse distances
from the cluster centre, recovering the galaxy velocity distributions
within them. We implemented a MCMC analysis, described in
Sec. \ref{fit}, to recover the shift of the mean of the velocity
distributions, which is proportional to the gravitational redshift
effect. We found a significant negative signal in all the four bins of
projected transverse distances from the cluster centre. Moreover, the
signal becomes more negative as the distance from the centre
increases, as expected.  We recovered an integrated gravitational
redshift signal of $\bar{\Delta}_{int} = -11.4 \pm 3.3$ km s$^{-1}$ up
to a distance of about $3$ Mpc from the cluster centre. This value is
in agreement with the expected value of approximately $-10$ km
s$^{-1}$, predicted in GR for clusters in the same range of masses as
the ones considered here.  The error on this integrated signal is
about $30\%$ lower with respect to what found in the previous works by
\citet{Sadeh_2015} and \citet{gravfromspider}.

We computed the theoretical gravitational redshift effect in three
different gravity theories: GR, $f(R)$ and sDGP. The gravitational
redshift model predictions are shown in
Fig. \ref{fig:deltapredctirad}. We compared our measurements with
theoretical predictions as a function of the transverse distance from
the cluster centre. This comparison is shown in
Fig. \ref{fig:whln>2misuram>14.2}. We implemented a new statistical
analysis method in order to discriminate among the different gravity
theories, as described in Sec. \ref{measures}. The free parameter of
this analysis is $\alpha$, which models the gravitational acceleration
in different gravity theories (by construction, $\alpha$ is equal to
unity in GR).  We obtained $\alpha = 0.86 \pm 0.25$. This result is in
agreement with GR and sDGP predictions, within the errors, while
marginally inconsistent with the $f(R)$ strong field model at
about $2\sigma$ significance, in line with literature results
  \citep[e.g.][]{Terukina_2014, Wilcox_2015}.

This work demonstrates that the peculiar velocity distribution of the
cluster member galaxies provides a powerful tool to directly
investigate the gravitational potentials within galaxy clusters and to
impose new constraints on the gravity theory on the megaparsec
scales. Further investigations are necessary to corroborate the
measurement method by exploiting cosmological simulations, especially
at high redshifts, and to improve the modelling for both galaxy
velocity distributions and gravitational redshift theoretical
predictions. The model improvements are necessary to take into account
the BCG proper motions, and to relax the assumption of the cluster
spherical symmetry and the NFW density profile. Forecasting analyses
are needed to compute the required number of clusters and associated
member galaxies necessary to discriminate among different gravity
theories with a high statistical significance. It will be useful to
investigate the effects possibly caused by mixing data from different
spectroscopic surveys, which can be done to increase the available
statistics by jointly combining different data sets.
Furthermore, it will be crucial to accurately calibrate the
  richness-mass scaling relation in different gravity models, to
  minimize the related systematic biases in the likelihood analysis.

To perform even stronger test on GR it will be necessary to reduce the
measurement errors, which mainly depend on the number of cluster
member galaxies available with spectroscopic redshift
measurements. Large cluster and galaxy samples from upcoming missions
will be crucial. In particular, the ESA {\em Euclid}
mission\footnote{\href{http://www.euclid-ec.org}{http://www.euclid-ec.org}}
\citep{laureijs2011, amendola2018} will detect $\sim 2 \times 10^6$
galaxy clusters up to $z\sim 2$ with a spectroscopic identification of
the cluster member galaxies \citep[see e.g.][]{Sartoris_2016}. The
scientific exploitation of the \textit{Euclid} cluster catalogues will
be key to obtain definite constraints on the gravity theory from
gravitaional redshifts inside galaxy clusters.

\section*{ACKNOWLEDGEMENTS}

We thank the anonymous referee for the useful comments to
  improve the quality of the paper. We acknowledge the grants ASI
n.I/023/12/0 and ASI n.2018-23-HH.0. LM acknowledges support from the
grant PRIN-MIUR 2017 WSCC32. We acknowledge the use of
  computational resources from the parallel computing cluster of the
  Open Physics Hub\footnote{\href{https://site.unibo.it/openphysicshub/en}{https://site.unibo.it/openphysicshub/en}} at the Physics and Astronomy Department in Bologna.

%
%

\bibliographystyle{aa} 
\bibliography{file} 

\begin{appendix}

\section{Testing the systematic uncertainties in the analysis} \label{appendix_systematics}
In this Appendix we describe the tests we conducted to investigate the
effects on the gravitational redshift measurements of the various
selections on the cluster member galaxies. Moreover, we discuss how
our results change when we assume the BCG as the cluster centre, as
done in the past literature works by \citet{Wojtak_2011},
\citet{Jimeno_2015} and \citet{Sadeh_2015}.

\subsection{Assuming the BCG as the cluster centre}
\label{BCGresultwhl}

In this work, we have used the average galaxy positions and redshifts
to estimate the cluster centres. To investigate the impact of this
choice on the measurement results, we repeat the analysis by assuming
the BCG as the cluster centre, as done in past literature works. We
construct the cluster member catalogues by using the same selection
criteria described in Sec. \ref{searchclustmemb}. When we estimate the
cluster centre from the BCG, all the cluster member galaxies are
selected all over again. The new set of member galaxies in the outer
cluster region is different in this case, since galaxies near the
edges of our selection can be either included or excluded, depending
on the centre choice, according to the selection criteria described in
Sec. \ref{searchclustmemb}. We obtain a cluster member catalogue with
$\num{3065}$ clusters and $\num{46819}$ cluster member galaxies.

We construct the background-corrected phase-space diagrams as
described in Sec. \ref{phasespace}, in order to compare the results
with those described in Sec. \ref{measures}.  Then, we fit the galaxy
velocity distributions to retrieve the gravitational redshift signal
as a function of the transverse distance from the cluster centre.  We
also estimate the $\alpha$ parameter by fitting the measured
$\bar{\Delta}$, using the same fitting procedure described in
Sec. \ref{measures}.

\begin{figure}
    \centering
    \includegraphics[width=\hsize]{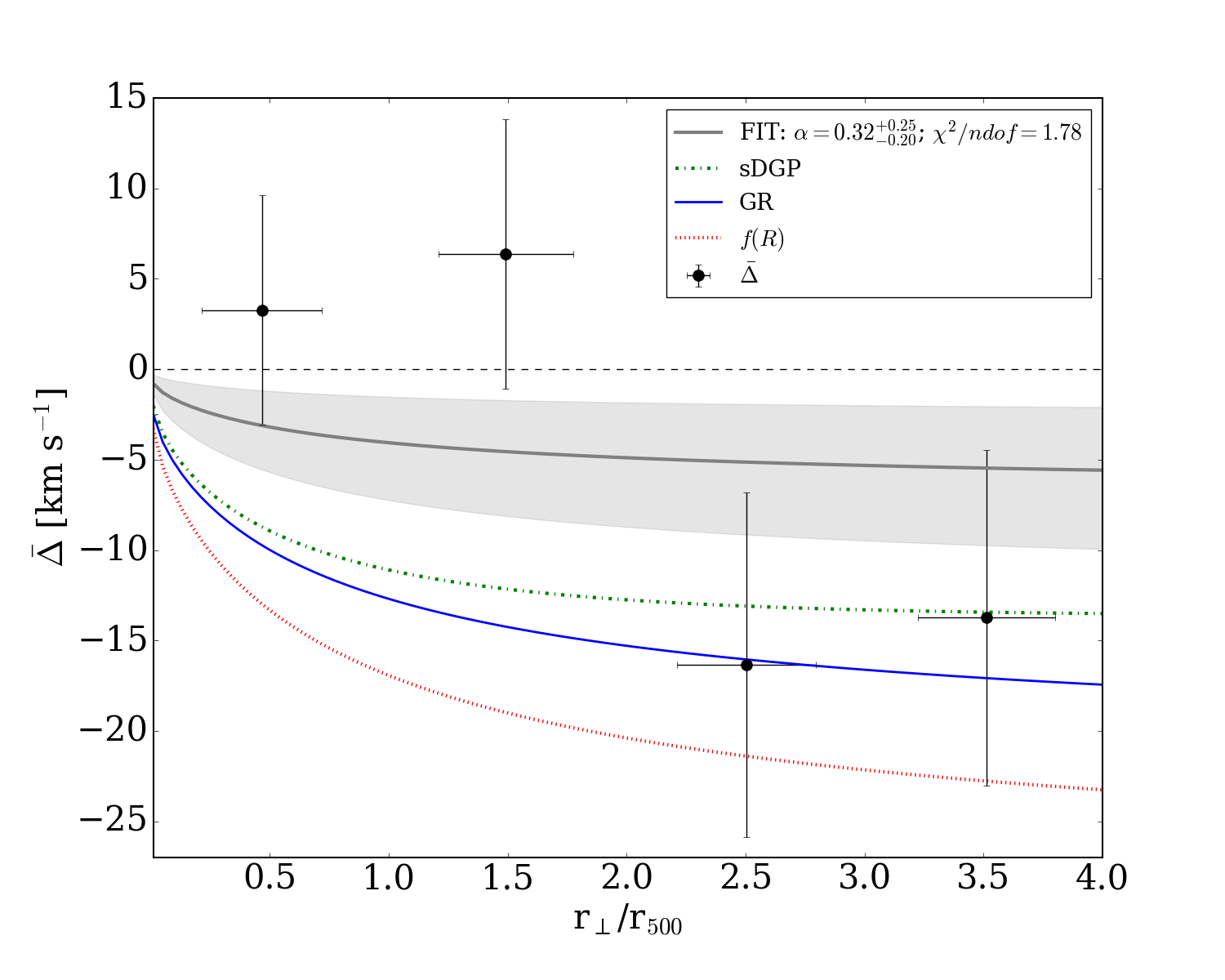}
    \caption{Best-fit model of $\bar{\Delta}$ from MCMC (grey solid
      line) for the analysed cluster member galaxies, which are
      identified assuming the BCG as the cluster centre. The symbols
      are the same as in Fig. \ref{fig:whln>2fit}.}
    \label{fig:whln>2fitbcg}
\end{figure}

Figure \ref{fig:whln>2fitbcg} shows the MCMC results, comparing the
estimated $\bar{\Delta}$ assuming the BCG as the cluster centre within
each distance bin and the theoretical predictions from GR, $f(R)$,
sDGP as functions of the transverse distance from the cluster centre.
Figure \ref{fig:whln>2fitbcg} shows that the measurements in the
outermost bins are in agreement with the theoretical models, while
those in the inner bins are not, showing positive values of the mean
of the galaxy velocity distribution. We interpret this result in the
inner bins as mainly caused by two effects. Firstly, the BCG peculiar
velocities causes a positive shift of $\bar{\Delta}$, as demonstrated
by \citet{kaiser2013}, which is similar to the TD effect, but less
intense. This effect was not included in the theoretical model,
described in Sec. \ref{model}, because it is expected to be a
second-order effect. Moreover, we do not have any information about
the BCG peculiar velocities. Further investigations are necessary to
understand the real impact of the BCG peculiar motions. Secondly, the
BCGs might be misidentificated, due to the surface brightness
modulation and velocity effects. In the case the true BCG is not
identified, we may erroneously consider as the centre a cluster member
galaxy which is instead a hot-population object, and may not lie near
the centre of the cluster gravitational potential well. Thus, the BCG
false identification might cause a positive shift of the mean of the
velocity distribution. A similar result was obtained by
\citet{Jimeno_2015} analysing the WHL12 cluster sample and assuming
the BCG as the cluster centre. \citet{Jimeno_2015} found marginally
positive values of $\bar{\Delta}$ in all the analysed bins of
trasverse distance from the cluster centre, up to $7$ Mpc.

In this case, we obtain $\alpha = 0.32_{-0.20}^{+0.25}$ with a reduced
$\chi^2$ equal to $1.78$. The best-fit model is marginally
inconsistent with any gravity theory we analysed, and it is almost
compatible with zero. The error on the $\alpha$ parameter is
asymmetric because we impose as a prior that $\alpha$ has to be
greater than zero (if $\alpha = 0$ we do not have a gravitational
force, while $\alpha<0$ would imply an anti-gravity force).

Hence, we tend to conclude that in this case we do not have a reliable
$\alpha$ estimation due to the BCG misidentification and peculiar
velocity effects, which are not taken into account in the
model. Further investigations are necessary to improve the modelling
when the BCG position is assumed as the cluster centre.

\subsection{Testing the selections}\label{testselectionwhl}

When we searched for the cluster member galaxies, we applied a number
of selections, that are described in Sec. \ref{searchclustmemb}. Here
we discuss the impact of these choices on the gravitational redshift
measurements.

\subsubsection{Selection of the cluster member galaxies}

In the analysis presented in this work, we considered only the
clusters which have at least $4$ associated member galaxies. We do
this following \citet{Wojtak_2011} and \citet{Jimeno_2015}, who made a
similar selection. This choice is useful to mitigate the problem of
false cluster identification. We test the effect of this selection by
measuring the gravitational redshift in the phase-space diagrams
constructed by changing the minimum number of cluster associated
member galaxies. We notice that if the minimum number increases above
$6$, the statistics becomes too low, and the measurement cannot be
performed due to the too small number of remaining clusters, which
increases Poisson noise in the velocity distributions. Considering a
number of members in the range between $3$ and $6$, the final results
are not significantly affected and remain in agreement with those
described in Sec. \ref{measures}. On the other hand, if the minimum
number of cluster member galaxies is less than $3$, the cluster false
identification significantly affects the measurement causing a
positive shift of the velocity distribution mean.

\subsubsection{Minimum number of galaxies used to compute the centre}

A second choice we made in our analysis was to select only the
clusters whose centres can be computed with at least $3$ member
galaxies. We made this choice to select only clusters whose average
redshifts have an error which is reduced by at least $50\%$ with
respect to the BCG spectroscopic redshift error. In fact, when we
compute the velocity distributions, the centre redshift error
propagates to all the redshifts of the cluster member galaxies. This
is a major issue, especially for the clusters with a large number of
members. We made several tests by arbitrarily increasing the redshift
error of the cluster centre, up to $5 \times 10^{-3}$. We notice that,
as the error increases, the velocity distributions have an increasing
larger positive shift of the mean. Further analyses are necessary to
investigate this effect. Moreover, we notice that if the number of
member galaxies used to estimate the centres increases above $5$, the
statistics becomes too low, which does not allow us to obtain any
sufficiently accurate measure. On the other hand, if this number is
less than $3$, the results are not statistically distinguishable from
the ones obtained considering the BCG as the cluster centre (these
measurements are described in Sec. \ref{BCGresultwhl}). For clusters
whose average centre positions are computed with $3-5$ galaxies, the
final results do not vary significantly from those described in
Sec. \ref{measures}.

\subsubsection{Cluster redshift range}

In our analysis, we selected only the clusters which have a redshift
smaller than $0.5$, to mitigate the problem of false cluster
identifications, and to mitigate the possible impact of the assumed
cosmological model on the measurements. In order to test this
selection, we measure the gravitational redshift changing the cluster
redshift cut-off. When we consider the low-redshift clusters with
$z<0.2$, the lack of statistics prevents us to obtain any
measurement. We also test the analysis up to $z = 0.6$. In this case
the final measurements do not vary significantly from those described
in Sec. \ref{measures}. Further studies are necessary to investigate
the method at higher redshifts, by exploiting cosmological
simulations, to quantify how gravitational redshift theoretical
predictions are affected by the redshift dependence of cosmological
parameters.

\subsubsection{Mass selection}\label{appmasssel}

A further selection we applied in our analysis was to consider only
the clusters which have masses above $1.5 \times 10^{14}$ M$_{\odot}$,
in order to mitigate the problem of the false cluster identification,
as described in Sec. \ref{searchclustmemb}. To investigate the impact
of including also lower mass clusters, we measure the gravitational
redshift as a function of the cluster mass. We split the cluster
member catalogue in four sub-samples covering different cluster mass
ranges. We do not take into account clusters with masses lower than $3
\times 10^{13}$ M$_{\odot}$, because for those the richness-mass
relation is not calibrated, as described in
Sec. \ref{whlclustcat}. For each sub-sample we construct the
background-corrected phase-space diagram and measure the integrated
gravitational redshift $\bar{\Delta}_{int}$ signal up to a transverse
distance from the cluster centre of $4 r_{500}$.

\begin{figure}
    \centering
    \includegraphics[width=\hsize]{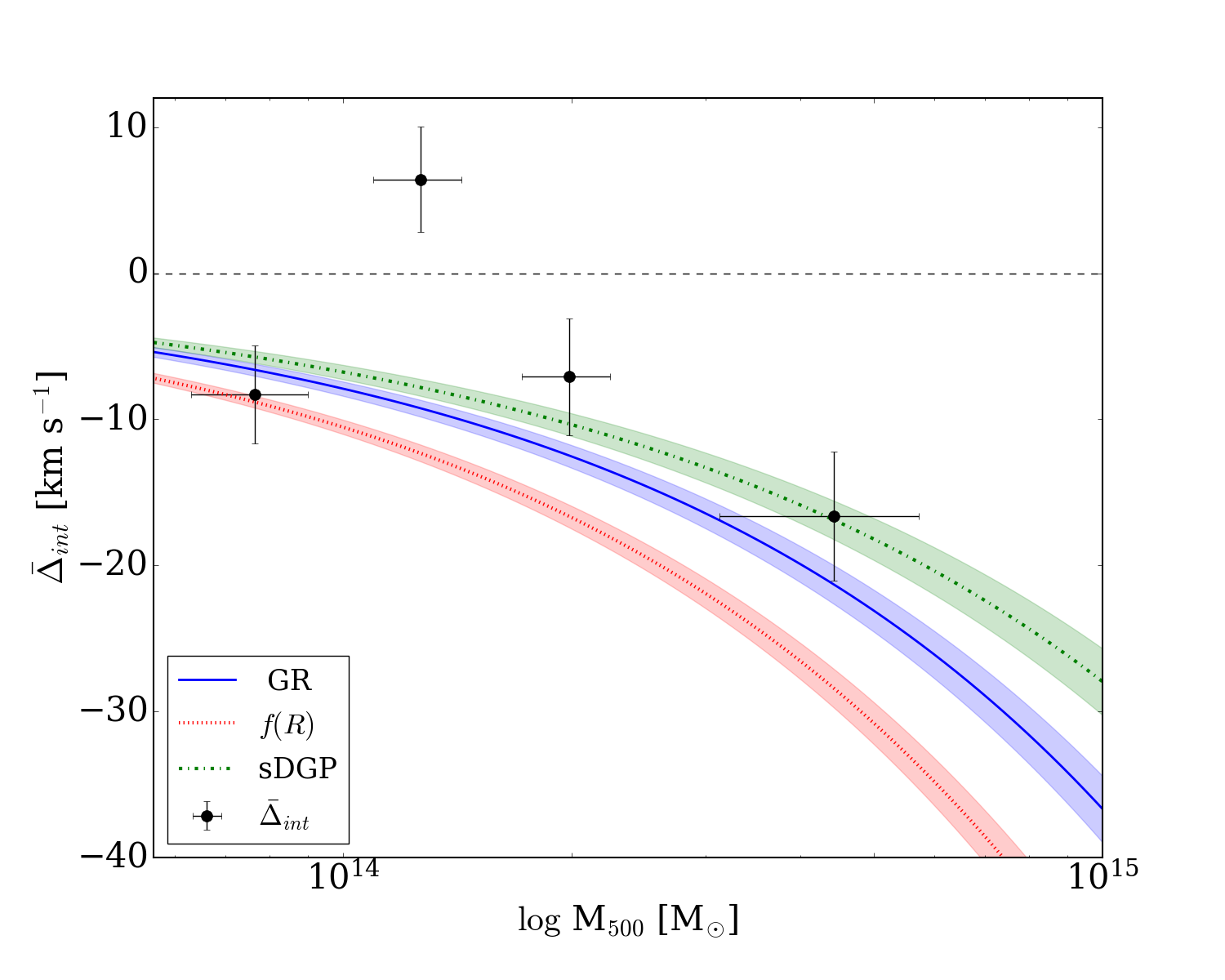}
    \caption{Comparison between the estimated integrated
        gravitational redshift signal $\bar{\Delta}_{int}$ up to $4
        r_{500}$ and the theoretical predictions from GR (blue solid
        line), $f(R)$ (red dotted line), sDGP (green dash-dotted line)
        as a function of the cluster mass. The shaded coloured areas
      show the model errors, while the black points show the estimated
      $\bar{\Delta}_{int}$. The vertical error bars represent the
      range of the $\bar{\Delta}_{int}$ parameter containing $68 \%$
      of the marginalised posterior probability, while the horizontal
      error bars show the dispersion of the cluster masses in a given
      bin.}
    \label{fig:whln>2misuramass}
\end{figure}

Figure \ref{fig:whln>2misuramass} shows the comparison between the
estimated integrated signal $\bar{\Delta}_{int}$ up to $4 r_{500}$,
within each sub-sample, and the GR, $f(R)$ and sDGP predictions. The
figure shows that the measurements at M$_{500} \gtrsim 2 \times
10^{14}$ M$_{\odot}$ are in agreement, within the errors, with the GR
and sDGP theoretical predictions. Moreover, in this mass range the
integrated signal up to $4 r_{500}$ becomes more negative as the
cluster mass increases, as expected. On the other hand, the
$\bar{\Delta}_{int}$ value for clusters with average mass of about
$1.5 \times 10^{14}$ M$_{\odot}$ is positive and it is not in
agreement with any prediction, while the $\bar{\Delta}_{int}$
measurement in the lowest mass range is again in agreement with all
theoretical predictions. The positive value of $\bar{\Delta}_{int}$ is
probably caused by a high percentage of false identified clusters in
this mass range.

To investigate the impact of the mass selection on the measurements,
we stack all the clusters with masses above $3 \times 10^{13}$
M$_{\odot}$ in a single background-corrected phase-space diagram, and
measure the gravitational redshift effect as a function of the
transverse distance from the cluster centre. We split the phase-space
diagrams in four bins of width equal to $r_{500}$, as done in
Sec. \ref{measures}. We also fit the measured
  $\bar{\Delta}$ with the procedure described in Sec. \ref{measures},
  to constrain the $\alpha$ parameter.

\begin{figure}
    \centering
    \includegraphics[width=\hsize]{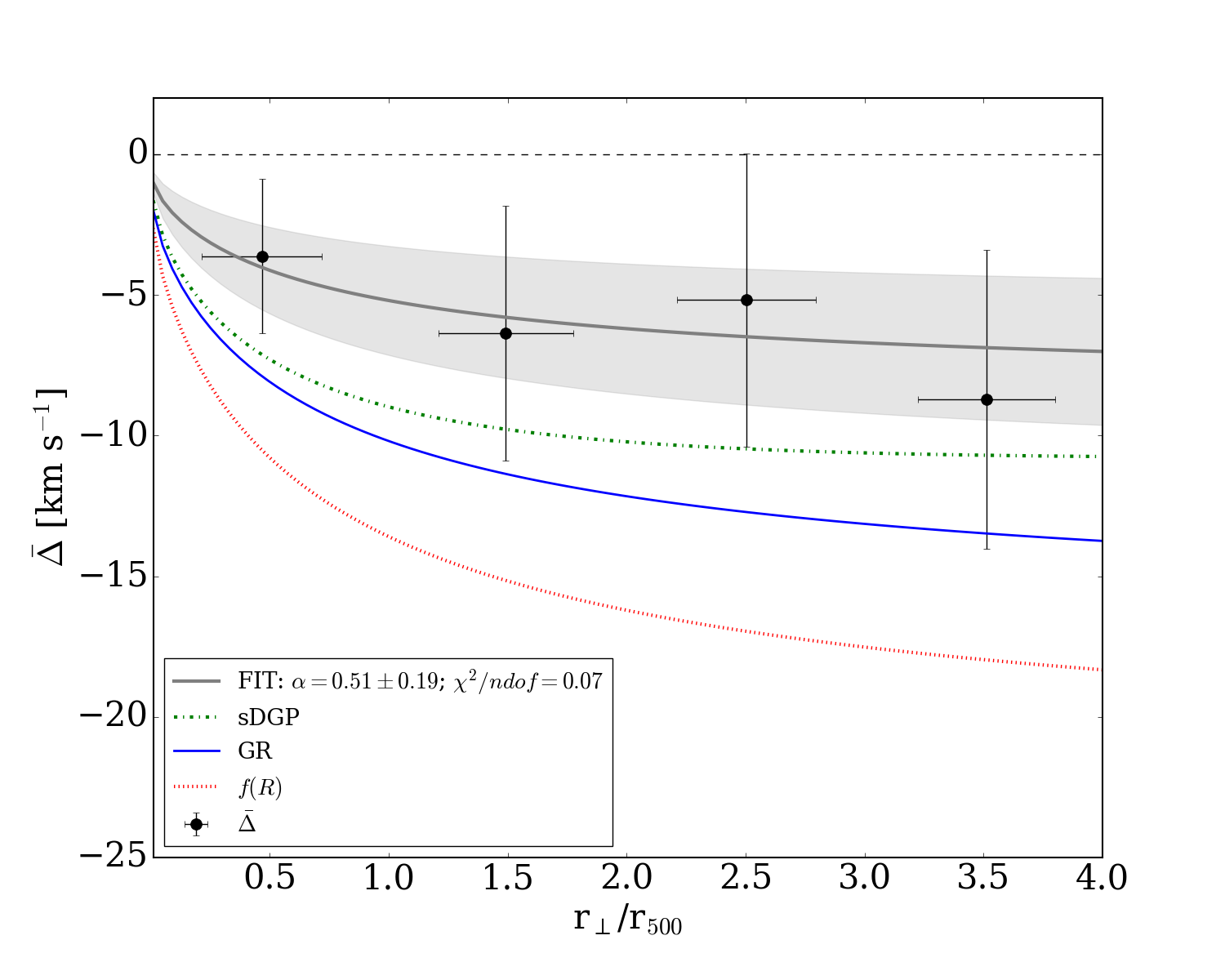}
    \caption{Best-fit model of $\bar{\Delta}$ from MCMC
      (grey solid line) for all the clusters with mass above $3 \times
      10^{13}$ M$_{\odot}$. The shaded grey area shows the $68 \%$
      uncertainty on the posterior median. The theoretical predictions
      as functions of the transverse distance from the cluster centre
      from GR (blue solid line), $f(R)$ (red dotted line), sDGP
      (green dash-dotted line) are shown for comparison.}
    \label{fig:whln>2fit}
\end{figure}

Figure \ref{fig:whln>2fit} shows the result of the MCMC and the
comparison between the estimated $\bar{\Delta}$ within each bin and
the GR, $f(R)$ and sDGP theoretical predictions.  Figure
\ref{fig:whln>2fit} shows that the measurements are in marginal
agreement only with the sDGP predictions in all the four bins,
differently from the mass-selected cluster member sample considered in
the analysis of this work (see Fig. \ref{fig:whln>2misuram>14.2}). We
interpret this result as possibly caused by the false identified
low-mass clusters. In this case we obtain a value of $\alpha$ equal to
$0.51 \pm 0.19$ with a reduced $\chi^2$ of $0.07$. This value is
marginally inconsistent with the estimation obtained from the mass
selected cluster member sample, which is $0.86 \pm 0.25$ (see
Fig. \ref{fig:whln>2fitm>14.2}).

\subsection{Cosmological dependence of the relation between the cluster masses and
observable proxies}\label{FRresult}
\begin{figure}
    \centering
    \includegraphics[width=\hsize]{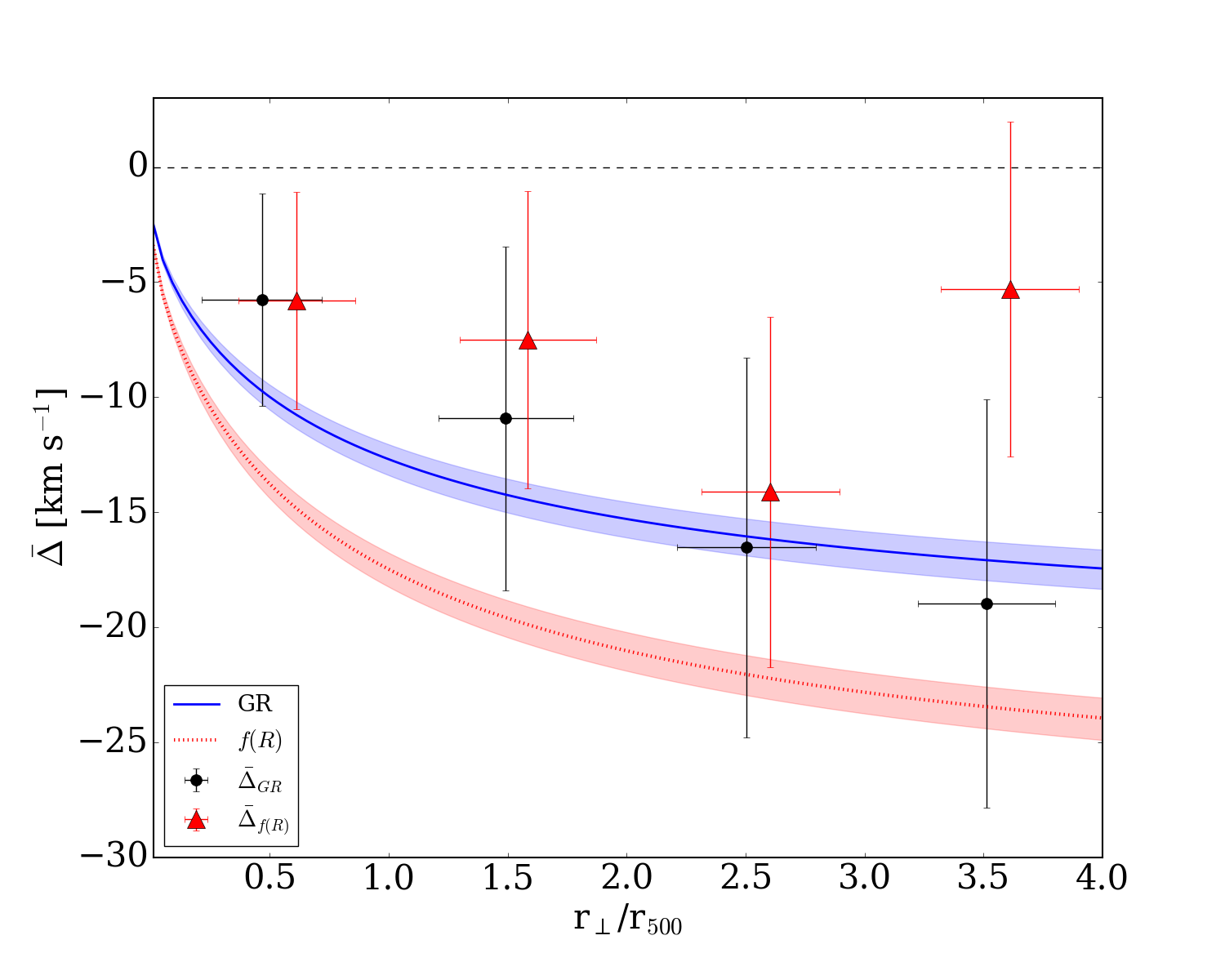}
    \caption{Comparison between the $\bar{\Delta}$ values
        computed with the scaling relation estimated in the $f(R)$
        strong field scenario, $\bar{\Delta}_{f(R)}$ (red triangles),
        and the ones assuming the reference scaling relation
        considered in this paper, $\bar{\Delta}_{GR}$ (black
        points). The former values are shifted horizontally by $0.1
        r_{\perp}/r_{500}$, for visual purposes. The theoretical
        $f(R)$ predictions are computed using the $M_{500}$ values
        obtained from the $f(R)$ scaling relation.  The shaded
        coloured areas show the theoretical uncertainties caused by
        the fitting uncertainties on the cluster mass distribution,
        and the dispersion of the cluster redshifts. The vertical
        error bars represent the range of $\bar{\Delta}$ parameter
        containing the $68 \%$ of the marginalised posterior
        probability, while the horizontal error bars show the
        dispersion of the galaxy transverse distances in each given
        bin.}
    \label{fig:FRresults}
\end{figure}

As described in Sec. \ref{measures}, we test the impact of
  modifying the relation between the cluster masses and observable
  proxies, when we assume $f(R)$ gravity theory.  We compute the new
  $M_{500}$ masses for each cluster in the $f(R)$ strong field
  scenario following \citet{Mitchell_2021}. For each cluster we
  compute also the radius, $r_{500}$, and concentration parameter,
  $c_{500}$. Then we perform again the full statistical analysis.
  Figure \ref{fig:FRresults} shows the comparison between
  $\bar{\Delta}_{f(R)}$, which are the results of this test, and
  $\bar{\Delta}_{GR}$, that is the measurements also shown in
  Figs. \ref{fig:whln>2misuram>14.2} and
  \ref{fig:whln>2fitm>14.2}. The $f(R)$ theoretical predictions are
  computed considering the new $M_{500}$ values. As shown in the
  Figure, the $\bar{\Delta}_{f(R)}$ measurements are shifted towards
  positive values with respect to $\bar{\Delta}_{GR}$, as discussed in
  Sec. \ref{measures}.

\end{appendix}

\end{document}